\documentclass[preprint]{aastex}

\usepackage{latexsym}

\begin{document}

\title{A Beginner's Guide to the Theory of CMB
  Temperature and Polarization Power Spectra in the Line-of-Sight Formalism}

\author{Yen-Ting Lin\altaffilmark{1,2} and Benjamin D. Wandelt\altaffilmark{1,3,4}}

\altaffiltext{1}{Department of Astronomy, University of Illinois,
Urbana, IL 61801}
\altaffiltext{2}{Current address: Department of Astrophysical Sciences, 
Princeton University, Princeton, NJ 08544; ytlin@astro.princeton.edu; also 
Department of Astronomy and Astrophysics, Pontificia Universidad Catolica de 
Chile}
\altaffiltext{3}{Department of Physics, University of Illinois,
Urbana, IL 61801; bwandelt@astro.uiuc.edu}
\altaffiltext{4}{NCSA Faculty Fellow}

\begin{abstract}
We present here a detailed, self--contained treatment of the
mathematical formalism for describing the theory of polarized
anisotropy in the cosmic microwave background. This didactic review is
aimed at researchers who are new to the field. We first develop
the mathematical tools for describing polarized scattering of CMB photons. Then 
we
take the reader through a detailed derivation of the line-of-sight formalism,
explaining the calculation of both temperature and polarization power spectra 
due to the scalar and tensor perturbations in a flat Universe.
\end{abstract}

\keywords{cosmology: theory -- cosmic microwave background -- polarization}

%%%%%%%%%%%%%%%%%%%%%%%%%%%%%%%%%%%%%%%%%%%%%%%%%%

\section{Introduction and Brief History of CMB Polarization}

In the gravitational instability paradigm of structure formation,  density
perturbations in the early Universe grow into the large scale structure we
observe today. The presence of the perturbations at the epoch of recombination
causes the
Cosmic Microwave Background (CMB) to be polarized through  Thomson
scattering. The degree of polarization in the CMB was predicted to be weak
(about $10\%$ or less than the temperature anisotropy at small angular scale,
even smaller at large angular scales, see e.g., \citealt{k96}),
the recent detection of polarization in the CMB by the {\it Degree Angular
Scale Interferometer}
(DASI, \citealt{dasi}, see also \citealt{leitch04}) and the measurement of the large-angle power spectrum
of correlations between temperature and polarization anisotropies by
{\it Wilkinson Microwave Anisotropy Probe} \citep[WMAP,][]{wmap,wmappol} were 
important confirmations of the paradigm \citep[e.g.][and references therein]{hu02}. 

There are several other reasons why studies of CMB polarization are important.
First, although studies of the temperature anisotropies in the CMB have
provided strong constraints on many of the cosmological parameters,
degeneracies between some combinations of parameters exist
\citep[e.g.,][]{sel97}. In principle, observation of polarization can help break some of
the degeneracies, such as that between
the overall amplitude of the temperature power spectrum and the epoch and degree
of reionization \citep{hu97a}. Second, some physical mechanisms
(e.g., gravity waves) only
contribute to very large angular scales, where our ability of extracting
cosmological information is limited by cosmic
variance. The additional information in the polarization anisotropies
can add valuable information for studies of cosmological physics on
super-large scales
\citep[e.g.,][]{zs97}. Related to this
point, we note that different perturbations modes (scalar, vector and
tensor) give distinguishable polarization patterns \citep{hu97a}.
Thus, from the construction of polarization power spectra we ought to be
able to investigate the nature and origin of perturbations presented in the
early
Universe.
 %To be more specific, inflation models generally predict existence of
%scalar and tensor perturbations, while defect models
%predict equivalent amount of vector and tensor perturbations.
%Therefore, any detection of the magnetic--type polarization pattern (see \S~3)
%can help distinct these two kinds of models.
Finally, unlike temperature
anisotropies
which are affected by various physical effects that occur between the
last scattering 
surface and present, polarization provides us with direct probe of  the last
scattering surface. It therefore helps distinguish the contributions to the temperature
power spectrum from gravitational potential and peculiar velocities
\citep[e.g.,][]{kks97}.

Although the importance of the CMB polarization was recognized very early on
(e.g.~\citealt{rees68,caderni78,negroponte80,kaiser83}), formalisms that enable fast
and accurate computations of the polarization field on the whole sky
were not realized until late 1990s.
%The early work on CMB polarization  focused on small angular scales
%where the flat-sky approximation is valid. One 
One main reason is that, in applying the Stokes $Q$ and $U$ parameters
to describe the polarization field, a fixed coordinate frame is needed. 
%In the flat-sky approximation the choice of this frame is trivial  \citep[e.g.,][]{sel97}. 
If one models polarization maps on the whole sphere,
this requirement complicates the computation. One way to avoid this is
to expand the polarization field by tensor harmonics or their
relatives, the spin-weighted harmonics. This was first done by two groups 
\citep{zs97,kks97}.  In this review we follow the formalism developed by 
\citet[][hereafter ZS97]{zs97} to reduce
the ``spin weight'' of the polarization fields to zero using the so-called
spin raising and lowering operators. In this framework the calculation
of the polarization power spectra becomes greatly simplified and very similar to
that of the temperature  power spectrum. The method applies to all angular 
scales, and is therefore applicable to the analysis of  all--sky surveys.
Due to its calculational convenience the spin weight formalism is the
technique of choice for codes  solving the linearized cosmological
Boltzmann equations such as CMBFAST or CAMB
\footnote{see http://www.cmbfast.org/ or http://cosmologist.info}.

In this review we first provide a self--contained overview on the basic physics 
of polarization and their mathematical description. Then we detail the derivations of several important 
results of ZS97 (presented in their sections II to IV).
We mainly provide the mathematical details of the calculations
that lead to temperature and polarization power spectra. This is
intended to help build the mathematical toolkit for new workers in 
CMB theory. We assume some basic familiarity with electromagnetism. One
subsection introduces polarization in a quantum mechanical notation, however,
this is not required in order to understand  what follows. Tutorials on the 
physical interpretation and visualization of CMB polarization have been 
available for some time and are summarized in \S~5.

This review is organized as follows: in \S~2 we present some basic ingredients
needed to understand the generation of polarization at the last scattering 
surface, including the Stokes parameters, the spin--weighted spherical 
harmonics, and the Thomson scattering. In \S~3
we introduce some basic definitions for calculations of temperature and
polarization power spectra, then formally calculate the power spectra induced by
scalar and tensor perturbations. 
We briefly summarize our results in \S~4 and provide references for further
reading in \S~5.
In Appendix A we discuss the parity of the polarization
modes $E$ and $B$; in  Appendix B we describe the line-of-sight integral
solutions to the Boltzmann equation.

%%%%%%%%%%%%%%%%%%%%%%%%%%%%%%%%%%%%%%%%%%%%%%%%%%%%%%%%%%%%%%%%%%%%%
%%%%%%%%%%%%%%%%%%%%%%%%%%%%%%%%%%%%%%%%%%%%%%%%%%%%%%%%%%%%%%%%%%%%%

\section{Polarization Basics}

In this section we discuss three topics which are helpful in understanding 
the physics of CMB polarization:
the Stokes parameters, the spin--weighted spherical harmonics, and  Thomson
scattering. In the discussion of the Stokes parameters, we present both a classical
(\S~\ref{sec:classical}) and quantum mechanical (\S~\ref{sec:qm}) picture for
describing electromagnetic radiation, then consider  observations with
finite bandwidth of radiation from uncoherent sources (\S~\ref{sec:prac}).
In our section of  spin--weighted spherical harmonics, we simply state the basic
features of this family of functions without proof.
Finally, the discussion on Thomson scattering  focuses first on the scattering
matrix, then on the generation of polarization at the last scattering surface.

We follow closely the discussions in \citet{shu}, \citet{chandra}, and \citet{rl} in sections
\ref{sec:classical}, \ref{sec:prac} and \ref{sec:scam}; the material presented
in sections \ref{sec:qm} and \ref{sec:lss} follows the discussions in 
\citet{k96,k99}; \S~\ref{sec:yslm} is based  on the appendix of ZS97, which
in turn is
drawn from \citet{np} and \citet{goldberg}. Note the difference in
the sign convention for rotation between that of ZS97 and the original 
reference; it is 
chosen so to match the practical convention used in the astronomical
literature. As far as possible we try to stick to a single convention in this
review, pointing out where our references differ along the way.

%%%%%%%%%%%%%%%%%%%%%%%%%%%%%%%%%%%%%%%%%%%%%%%%%%
\subsection{Stokes Parameters}

%%%%%%%%%%%%%%%%%%%%%%%%%%%%%%%%%%%%%%%%%%%%%%%%%%
\subsubsection{Classical Description
\label{sec:classical}}

Consider a  plane electromagnetic wave propagating along the
$\hat{z}$ direction. Its Fourier decomposition can be expressed as
\[
{\bf E}(z,t) = \int_{-\infty}^\infty (\hat{x} \mathcal{E}_x {\rm e}^{i \phi_x} +
    \hat{y} \mathcal{E}_y {\rm e}^{i \phi_y} ) {\rm e}^{i (k z- \omega t)}
    d k,
\]
where $\mathcal{E}$ and $\phi$ are real quantities denoting the amplitudes and phases in
the two transverse directions marked by the unit vectors $\hat{x}$ and
$\hat{y}$. The angular frequency of the wave is $\omega = k c$.
The real part of a given ${\bf k}$ mode is
\begin{equation}
{\bf E}_k = \hat{x} \mathcal{E}_x \cos(k z- \omega t+\phi_x)
    + \hat{y} \mathcal{E}_y \cos(k z- \omega t+\phi_y).
\label{eq:ek1}
\end{equation}
On the $x-y$ plane, the tip of the electric vector ${\bf E}_k$ will trace out an
ellipse as a function of time. Let the angle between the major axis of the ellipse ($x'$)
and $x$--axis be $\chi$, i.e.,
\begin{equation}
\left( \begin{array}{c}
\hat{x}' \\
\hat{y}'
\end{array} \right) =
\left( \begin{array}{cc}
\cos\chi & \sin\chi \\
-\sin\chi & \cos\chi
\end{array} \right)\
\left( \begin{array}{c}
\hat{x} \\
\hat{y}
\end{array} \right).
\label{eq:rotmatrix}
\end{equation}
We can choose the zero point of time so that ${\bf E}_k(t=0)$ is
purely in the 
$\hat{x}'$ direction:
\begin{equation}
{\bf E}_k(t=0) = \hat{x}' E_1 \cos\omega t + \hat{y}' E_2 \sin\omega t,
\label{eq:ek2}
\end{equation}
where
\begin{equation}
E_1^2 + E_2^2 = \mathcal{E}_x^2 + \mathcal{E}_y^2 \equiv \mathcal{E}_0^2
\label{eq:amp1}
\end{equation}
and
\begin{equation}
E_1 = \mathcal{E}_0 \cos\beta,\ \ \ \ \ E_2 = \mathcal{E}_0 \sin\beta,\ \ \ (-\pi/2 \le \beta \le \pi/2).
\label{eq:amp2}
\end{equation}
The ellipticity angle $\beta$ determines the shape of the ellipse. For example, for 
$\beta = \pm \pi/4$, the ellipse becomes a circle, and the wave is circularly 
polarized; if $\beta = 0, \pm \pi/2$, the ellipse ``collapses'' into a line, and
the wave is linearly polarized (also see below).

The quantities in Eqns~(\ref{eq:ek1}) and (\ref{eq:ek2}) can be related with the
help of Eqns~(\ref{eq:rotmatrix}) and (\ref{eq:amp2}); the coefficients of
$\hat{x} \cos \omega t$ give
$\mathcal{E}_x \cos \phi_x = E_1 \cos\chi = \mathcal{E}_0\cos\beta\cos\chi$,
those of $\hat{x} \sin \omega t$ give
$\mathcal{E}_x \sin \phi_x = -E_2 \sin\chi = -\mathcal{E}_0\sin\beta\sin\chi$,
those of $\hat{y} \cos \omega t$ give
$\mathcal{E}_y \cos \phi_y = E_1 \sin\chi = \mathcal{E}_0\cos\beta\sin\chi$,
those of $\hat{y} \sin \omega t$ give
$\mathcal{E}_y \sin \phi_y = E_2 \cos\chi = \mathcal{E}_0\sin\beta\cos\chi$.

The Stokes parameters are then defined as follows:
\begin{eqnarray}
\label{eq:i}
I & \equiv & \mathcal{E}_x^2 + \mathcal{E}_y^2  = \mathcal{E}_0^2,\\
\label{eq:q}
Q & \equiv & \mathcal{E}_x^2 - \mathcal{E}_y^2 = \mathcal{E}_0^2
    \cos2\beta \cos2\chi, \\
\label{eq:u}
U & \equiv & 2\mathcal{E}_x \mathcal{E}_y \cos(\phi_y-\phi_x)
    = \mathcal{E}_0^2 \cos2\beta \sin2\chi, \\
V & \equiv & 2\mathcal{E}_x \mathcal{E}_y \sin(\phi_y-\phi_x)
    = \mathcal{E}_0^2 \sin 2\beta.
\end{eqnarray}
Note that when the wave is monochromatic these parameters are related by
$I^2 = Q^2 + U^2 + V^2$.

The Stokes parameters defined in this way are all {\it real} quantities.
The $I$ parameter measures the radiation intensity and is positive. The other
parameters describe the polarization state and can take either positive or
negative sign.  The parameters
$Q$ and $U$ measure the orientation of the ellipse relative to the $x$-axis
and define the polarization angle
\begin{equation}
\chi \equiv \frac{1}{2} \tan^{-1} \frac{U}{Q}
\end{equation}
and the polarization vector
\begin{equation}
{\bf P} = (Q^2 + U^2)^{1/2} \hat{x}'.
\end{equation}
A few comments on the various sign conventions  of Stokes $Q$ and $U$ are in order.
In the usual spherical coordinates, $Q >0$ for a N-S (longitudinal,
$\pm \hat{{\rm e}}_{\theta}$) polarization, $Q<0$ for
a E-W (azimuthal, $\pm \hat{{\rm e}}_{\phi}$) component; on the other hand,
a NE-SW ($\pm (\hat{{\rm e}}_{\phi}-\hat{{\rm e}}_{\theta})/\sqrt{2}$)
polarization component means a positive $U$ while a NW-SE
($\pm (\hat{{\rm e}}_{\phi}+\hat{{\rm e}}_{\theta})/\sqrt{2}$)
component means a negative one \citep{hu97a}.
The last parameter, $V$, measures the relative
strengths of two polarization states: {\it linear}
polarized light has $V=0$; light for which $Q = U = 0$ and $V>0\,(V<0)$ is
right-- (left--) {\it circularly} polarized. 
%This can also be seen from the
%$\beta$ dependence of $V$; $\beta$ measures the shape of the ellipse. When
%$\beta=0$, the ellipse ``collapses'' into a line, thus linear polarization.
In particular, for unpolarized, or ``natural'' light, $Q=U=V=0$.

It is apparent that, with respect to a rotation about the $\hat{z}$ axis,
$I$, $V$ and $P^2=Q^2+U^2$ are
invariant, because they are independent of $\chi$,  the
angle between the polarization vectors ($\hat{x}'\ \&\ \hat{y}'$) and the
artificially chosen coordinates ($\hat{x}\ \&\ \hat{y}$). However, $Q$ and
$U$ transform in the following way under a {\it counterclockwise}
rotation of the $x-y$ plane through an angle $\alpha$ about the
$\hat{z}$--axis, as can be obtained by letting $\chi \rightarrow \chi-\alpha$
in Eqns~(\ref{eq:q}) and (\ref{eq:u}):
\begin{eqnarray}
\nonumber
Q' & = & Q \cos2\alpha + U \sin2\alpha, \\
U' & = & - Q \sin2\alpha + U \cos2\alpha.
\label{eq:qutransform}
\end{eqnarray}
A compact way of writing these is through a combination of $Q$ and $U$:
$(Q\pm iU)' = {\rm exp}(\mp 2i\alpha) (Q \pm iU)$.
In fact this is the property that makes $(Q\pm iU)$ a  spin--$\pm$2 quantity,
a fact that we will exploit heavily in section \ref{sec:yslm}.

%%%%%%%%%%%%%%%%%%%%%%%%%%%%%%%%%%%%%%%%%%%%%%%%%%
\subsubsection{ Quantum Mechanical Description
\label{sec:qm}}

A quantum mechanical description of the Stokes parameters\footnote{A detailed
    understanding of this subsection is not required to understand the
    remainder of this review - the reader unfamiliar with Dirac notation can
    skip to \ref{sec:prac} for the first reading.}, which is closely
related to the photon density matrix, can be found in \citet{k96}.
Consider two orthonormal linear polarization basis kets $|\epsilon_1\rangle$
and $|\epsilon_2\rangle$. An arbitrary state vector can be spanned by these
basis kets
\begin{equation}
|\epsilon\rangle = a_1{\rm e}^{i\delta_1} |\epsilon_1\rangle
    + a_2{\rm e}^{i\delta_2} |\epsilon_2\rangle.
\end{equation}
The Stokes parameters can be defined as the expectation values of four
operators
\begin{eqnarray}
\nonumber
\hat{I} & = & |\epsilon_1\rangle \langle \epsilon_1|+
    |\epsilon_2\rangle \langle \epsilon_2|, \\
\nonumber
\hat{Q} & = & |\epsilon_1\rangle \langle \epsilon_1|-
    |\epsilon_2\rangle \langle \epsilon_2|, \\
\nonumber
\hat{U} & = & |\epsilon_1\rangle \langle \epsilon_2|+
    |\epsilon_2\rangle \langle \epsilon_1|, \\
\label{eq:stokeoper}
\hat{V} & = & i|\epsilon_2\rangle \langle \epsilon_1|-
    i|\epsilon_1\rangle \langle \epsilon_2|.
\end{eqnarray}
For example, the expectation value of $\hat{I}$ is
\begin{eqnarray*}
I \equiv \langle \hat{I} \rangle = \langle \epsilon|\hat{I}|\epsilon\rangle
    = \left( a_1^* {\rm e}^{-i\delta_1}\langle \epsilon_1| +
    a_2^* {\rm e}^{-i\delta_2}\langle \epsilon_2| \right) \hat{I}
    \left( a_1{\rm e}^{i\delta_1} |\epsilon_1\rangle
    + a_2{\rm e}^{i\delta_2} |\epsilon_2\rangle \right)
    = |a_1|^2 + |a_2|^2,
\end{eqnarray*}
consistent with the definition in Eqn~(\ref{eq:i}).

Since polarization represents a mixture of the two degrees of freedom of a
photon, the photon density matrix has the components\citep[e.g.][]{sakurai} 
\begin{equation}
\hat{\rho} =
\left( \begin{array}{cc}
\rho_{11} & \rho_{12} \\
\rho_{21} & \rho_{22}
\end{array} \right).
\label{eq:rho}
\end{equation}
The components of $\rho$ can be expressed in terms of  expectation values of
the operators defined in Eqn~(\ref{eq:stokeoper}). We use the definition of
operator expectation values in terms
of the density matrix; e.g.~
%\begin{eqnarray*}
$Q \equiv \langle \hat{Q} \rangle = {\rm tr}(\hat{\rho}\hat{Q})
    = \rho_{11}-\rho_{22}.$
%\end{eqnarray*}
Similarly, $I = \rho_{11}+\rho_{22}$, $U \equiv \langle \hat{U} \rangle
= \rho_{12}+\rho_{21}$, and $V \equiv \langle \hat{V} \rangle
= i(\rho_{12}-\rho_{21})$. From these we can express the density matrix
in terms of the Stokes parameters and the Pauli matrices $\sigma_i$:
\begin{eqnarray}
\label{eq:phomatrix}
\nonumber
\hat{\rho} & = & {1 \over 2}
\left( \begin{array}{cc}
I+Q & U-iV \\
U+iV & I-Q
\end{array} \right) \\
  & = & {1\over 2} \left( I {\bf 1} +U \sigma_1 + V \sigma_2 +Q \sigma_3
    \right),
\end{eqnarray}
where ${\bf 1}$ is the $2\times2$ unit matrix.

The density matrix defined here is equivalent to the intensity tensor
defined in ZS97: the difference is that the intensity tensor is described
in terms of fractional change in the CMB temperature (recall that the
intensity is proportional to the fourth power of temperature, $I \propto T^4$).

%%%%%%%%%%%%%%%%%%%%%%%%%%%%%%%%%%%%%%%%%%%%%%%%%%
\subsubsection{Practical Description
\label{sec:prac}}

So far we have only considered monochromatic waves. In reality,
measurements of electromagnetic waves usually actually measure averages over
several oscillation periods and over a range of frequencies. We thus define
the averaged Stokes parameters as
\begin{eqnarray}
\nonumber
\overline{I} & \equiv & \left<\mathcal{E}_x^2 + \mathcal{E}_y^2 \right>
    = \left< \mathcal{E}_0^2 \right>,\\
\nonumber
\overline{Q} & \equiv & \left<\mathcal{E}_x^2 - \mathcal{E}_y^2 \right>=
        \overline{I} \cos2\beta \cos2\chi, \\
\nonumber
\overline{U} & \equiv & 2\left<\mathcal{E}_x \mathcal{E}_y \right>\cos(\phi_y-\phi_x)
    = \overline{I} \cos2\beta \sin2\chi, \\
\overline{V} & \equiv & 2\left<\mathcal{E}_x \mathcal{E}_y \right>\sin(\phi_y-\phi_x)
    = \overline{I} \sin 2\beta.
\end{eqnarray}
Note that the time scale for variations in $\mathcal{E}_i(t)$ and $\phi_i(t)$
($i=x,y$) must be large compared to the period of the wave.

An important property of the Stoles parameter is that they are additive for
incoherent superpositions of waves. Light from astronomical sources is therefore
not expected to be completely elliptically polarized, for they
generally come from different regions of the source, with different
polarization amplitude, vector, and phases. Suppose we receive a beam composed
of a mixture of independent, elliptically polarized light,
${\bf E}_k = \sum_j {\bf E}_k^j$, where different ${\bf E}_k^j$ components do
not possess coherent phases with each other.
Upon averaging over time and bandwidth, we have
\begin{eqnarray*}
\overline{I} = \sum_j \overline{I}^j,\ \ \ \  \overline{Q} = \sum_j \overline{Q}^j,\ \ \ \
\overline{U} = \sum_j \overline{U}^j,\ \ \ \  \overline{V} = \sum_j \overline{V}^j.
\end{eqnarray*}
In this case, it can be shown that
$\overline{I}^2 \ge \overline{Q}^2+ \overline{U}^2+ \overline{V}^2$ 
(\citealt{chandra}, p.32; \citealt{rl}, p.67). Because of this
property, it is always possible to decompose the observed radiation
$(\overline{I}, \overline{Q}, \overline{U}, \overline{V})$ into two
components: one completely unpolarized, the other elliptically polarized.
Specifically, for the polarized part,
$(\overline{I}_p, \overline{Q}_p, \overline{U}_p, \overline{V}_p) =
((\overline{Q}^2+ \overline{U}^2+\overline{V}^2)^{1/2}, \overline{Q}, \overline{U}, \overline{V})$;
for the unpolarized part, $(\overline{I}_u, \overline{Q}_u, \overline{U}_u, \overline{V}_u) =
(\overline{I}-\overline{I}_p, 0, 0, 0)$. Finally, the fractional polarization is
given by the ratio $\overline{I}_p/\overline{I}$. In what follows we will only focus
on the polarized part of the radiation, and ignore the subscript $p$.

%%%%%%%%%%%%%%%%%%%%%%%%%%%%%%%%%%%%%%%%%%%%%%%%%%
\subsection{Spin--Weighted Spherical Harmonics
\label{sec:yslm}}

Having described the properties of polarized light, we now develop the
mathematical machinery necessary to represent angular distribution of the polarization of the
CMB on the celestial sphere. The representation that
is most convenient for making contact with cosmological theory is in terms of
spin-weighted spherical harmonics.

A quantity $\eta$ that transforms as $\eta' = {\rm exp}(-s\, i\alpha) \eta$
under a rotation of an angle $\alpha$ is defined as of {\it spin--weight s}.
Examples we have already encountered are: the scalar temperature anisotropy,
which is a spin-$0$ quantity, and $Q \pm iU$ which is  a spin--$\pm$2 quantity.
It is important to note that the rotation here does \emph{not} mean a global rotation
that changes labeling of coordinates of the sphere (the sky). What is meant is
simply a rotation in the plane tangent to the point of interest on the
sphere, ie., a rotation of  the coordinates defined on that tangent plane.

Just as it is the case that a scalar (spin-$0$) function  on a sphere can be expanded into
a series of spherical harmonics $Y_{lm}(\theta,\phi)$ (also
spin-$0$ functions), a spin--$s$ function can be
expanded in spin--$s$ spherical harmonics $_sY_{lm}(\theta,\phi)$. For each
$|s| \le l$, they form a complete, orthogonal basis on the sphere:
\begin{equation}
\int d\Omega\, _sY^*_{l'm'}(\theta,\phi) _sY_{lm}(\theta,\phi) = \delta_{l'l}
    \,\delta_{m'm},
\end{equation}
\begin{equation}
\sum_{l,m}\, _sY_{lm}^*(\theta,\phi) _sY_{lm}(\theta',\phi')
    = \delta(\phi-\phi')\,\delta(\cos\theta-\cos\theta').
\end{equation}

Spin--weighted spherical harmonics were devised by \citet{np}. They have been
applied mostly in theories of multipole expansion of gravitational waves.
In this section we follow the discussion in ZS97, simply state some
of their properties without proof. See \citet{np,goldberg,pr,hu97b} for detailed
descriptions of the function. Also see \citet{kip,kks97} for relations between
this family of functions with other tensor spherical harmonics.

There exist a pair of operators, $\eth$ and $\bar{\eth}$, known as the spin
raising and lowering operators, respectively (in this review, for notational
convenience, they are denoted as:
$\sharp \equiv \eth$, $\flat \equiv \bar{\eth}$).
These operators have the property of raising/lowering the spin weight of a
function; denoting a quantity in a frame rotated $\psi$ from the original
frame by a prime, we have $(\sharp\,_sf)' = {\rm exp}(-i (s+1)\psi)\,
(\sharp\,_sf)$, and $(\flat\,_sf)' = {\rm exp}(-i (s-1)\psi)\, (\flat\,_sf)$.
Their explicit form is
\begin{eqnarray}
\sharp\,_sf(\theta,\phi) & = & -\sin^s\theta\,\left[ \partial_\theta
    + {i\over \sin\theta} \partial_\phi \right] \sin^{-s}\theta\,
    _sf(\theta,\phi),\\
\flat\,_sf(\theta,\phi) & = & -\sin^{-s}\theta\,\left[ \partial_\theta
    - {i\over \sin\theta} \partial_\phi \right] \sin^{s}\theta\,
    _sf(\theta,\phi).
\label{eq:sandf}
\end{eqnarray}

Suppose we have spin--$\pm2$ functions $_{\pm2}f(\theta,\phi)$ whose 
$\phi$--dependence satisfies $\partial_\phi\,_sf=im\,_sf$. Acting twice with the
spin raising/lowering operators on $_{\pm2}f$ gives (ZS97)
\begin{eqnarray}
\label{eq:f2}
\flat^2\,_2f(\mu,\phi) & = & \left( -\partial_\mu+{m\over 1-\mu^2} \right)^2
    \left[ (1-\mu^2)\,_2f(\mu,\phi) \right], \\
\label{eq:s2}
\sharp^2\,_{-2}f(\mu,\phi) & = & \left( -\partial_\mu-{m\over 1-\mu^2} \right)^2
    \left[ (1-\mu^2)\,_{-2}f(\mu,\phi) \right],
\end{eqnarray}
where we  have used the  notation $\mu \equiv \cos\theta$.
Notice that, $\flat^2\,_2f$ means acting $\flat$ on a spin--2 function $_2f$ first,
then acting with another $\flat$ on the resulting spin--1 function $\flat\,_2f$; the
final result is a spin-$0$ function, which is invariant under a rotation.
The same applies to the case $\sharp^2\,_{-2}f$.

One can relate the spin--$s$ spherical harmonics the usual spherical harmonics
by \citep{goldberg}
\begin{equation}
_sY_{lm}  =  \left[ {(l-s)!\over (l+s)!} \right]^{1/2} \sharp^s\,Y_{lm}
\end{equation}
for $0\le s\le l$, and
\begin{equation}
_sY_{lm}  =  \left[ {(l+s)!\over (l-s)!} \right]^{1/2} (-1)^s \flat^{-s}\,Y_{lm}
\end{equation}
for $-l \le s\le 0$. Using these, we obtain \citep[e.g.][]{hu97b}
\[
_{\pm 2}Y_{lm} = \left[ {(l-2)!\over (l+2)!} \right]^{1/2} \left[ 
  \partial_\theta^2 - \cot \theta \partial_\theta \pm \frac{2i}{\sin \theta}
  (\partial_\theta - \cot \theta) \partial_\phi - \frac{1}{\sin^2 \theta} 
  \partial_\phi^2 \right] Y_{lm}.
\]
For completeness, the explicit expression for spin--$s$ spherical
harmonics is \citep{goldberg}
\begin{eqnarray}
\nonumber
_sY_{lm}(\theta,\phi) & = & {\rm e}^{im\phi}
  \sqrt{ {2l+1\over 4\pi} {(l+m)!(l-m)!\over(l+s)!(l-s)!}}\,\sin^{2l}(\theta/2) \\
 & & \times
  \sum_r \left( \begin{array}{c}
l-s \\
r
\end{array} \right) \left( \begin{array}{c}
l+s\\
s-m+r
\end{array} \right)(-1)^{l-s-r} \cot^{s-m+2r}(\theta/2),
\end{eqnarray}
where 
\begin{eqnarray}
\nonumber
\left( \begin{array}{c}
p \\
q
\end{array} \right) \equiv { p! \over (p-q)!\, q! }
\end{eqnarray}
are the binomial coefficients.

Useful properties of the spin--weighted spherical 
harmonics are \citep[e.g.][]{goldberg}:
\begin{eqnarray}
\nonumber
_sY_{lm}^* & = & (-1)^{m+s}\,_{-s}Y_{l-m}, \\
\nonumber
\sharp\,_sY_{lm} & = & \left[ (l-s)(l+s+1) \right]^{1/2}\,_{s+1}Y_{lm}, \\
\nonumber
\flat\,_sY_{lm} & = & -\left[ (l+s)(l-s+1) \right]^{1/2}\,_{s-1}Y_{lm}, \\
\flat\,\sharp\,_sY_{lm} & = & -(l-s)(l+s+1)\,_sY_{lm}.
\label{eq:additional}
\end{eqnarray}
These relations also fix our sign/phase convention.
Note that ZS97 choose a different sign convention for the spin weighted spherical harmonics, which results in the relation $_sY_{lm}^\ast=(-1)^{s}\ _sY_{l-m}$.

%%%%%%%%%%%%%%%%%%%%%%%%%%%%%%%%%%%%%%%%%%%%%%%%%%
\subsection{Thomson Scattering}

%%%%%%%%%%%%%%%%%%%%%%%%%%%%%%%%%%%%%%%%%%%%%%%%%%
Now that we have set the mathematical scene we are ready to begin the
description of the physics of CMB polarization.

\subsubsection{Scattering Matrix
\label{sec:scam}}

The process of scattering off a photon by a charged particle where
there is no change in photon energy is
called Rayleigh scattering. In particular, when the charged particle is
an electron, the process is known as Thomson scattering.

\begin{figure}[htb]
\epsscale{0.65}
\plotone{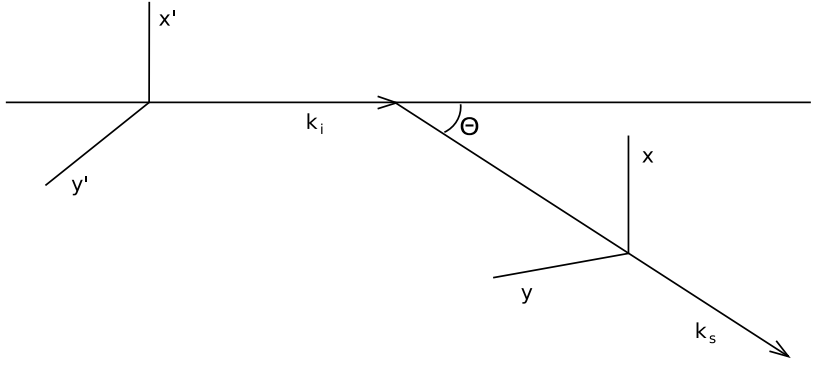}
\caption{Geometry of the Thomson scattering. Here $k_i$ and $k_s$ denote the 
    wave vectors of the incident and scattered radiation, respectively.
    }
\label{fig:geometry}
\end{figure}

Imagine an electron at the origin for instance just before the epoch of
recombination. This electron is accelerated by a  incoming plane wave of
radiation with wave vector  $\hat{k}_i$ and re-radiates an outgoing (scattered) wave along
$\hat{k}_s$. We will call the plane spanned by $\hat{k}_i$ and $\hat{k}_s$,
the \emph{scattering plane} (see Fig.~\ref{fig:geometry}).

If the incoming radiation is polarized parallel to the scattering plane (which
means that the electric field vector is in this plane), the differential cross 
section of Thomson scattering is (e.g., \citealt{rl})
\begin{equation}
\left. \frac{d\sigma}{d\Omega}\right|_{pol} = {3 \sigma_T \over 8\pi}
        | \hat{k}_i \cdot \hat{k}_s |^2,
\end{equation}
where $\sigma_T$ is the Thomson cross section, and the solid angle
$d\Omega=d(\cos\theta)d\phi$ defined in the usual spherical coordinates.  Note 
that the outgoing
radiation is still polarized parallel to the scattering plane, but for
right--angle scattering there is no 
outgoing radiation along  $\hat{k}_s$ since $\hat{k}_i \cdot \hat{k}_s =0$.

Next let us consider unpolarized radiation, which can be regarded as
an independent superposition of two linearly polarized waves with
perpendicular axes. We can choose one wave as polarized in the scattering
plane, the other  perpendicular to the plane. Then
the differential cross section, which is the sum of two polarized cross 
sections, is
\begin{equation}
\left. \frac{d\sigma}{d\Omega} \right|_{un} = {3 \sigma_T \over 16\pi}
        \left( | \hat{k}_i \cdot \hat{k}_s |^2 +1 \right);
\end{equation}
that is, the ratio of the intensities of the scattered light in directions
parallel and perpendicular to the scattering plane is $\cos^2\theta$, where
$\cos\theta = \hat{k}_i \cdot \hat{k}_s$. For right--angle
scattering, $\theta=\pi/2$, the scattered light
is completely linearly polarized in the direction perpendicular to the
scattering plane.

Formally, we denote the incident radiation as $\mathcal{I}' =
(I'_\parallel, I'_\perp, U', V')$, where $I'_\parallel$ and $I'_\perp$ are the
intensities parallel and perpendicular to the plane of scattering,
respectively. The total intensity is $I' = I'_\parallel+I'_\perp$, and
$Q' = I'_\perp-I'_\parallel$\ \footnote{The definition is somewhat arbitrary
(up to a sign change). Here we follow that of \citet{k96}.}.
The incident and scattered radiation is related by 
the scattering matrix (or ``phase matrix'', e.g., \citealt{chandra})
\begin{equation}
%{\bf R} =  {3\sigma_T\over 8\pi}
{\bf R} =  {3\over 8\pi}
\left( \begin{array}{cccc}
\cos^2\theta & 0 & 0 & 0 \\
0 &            1 & 0 & 0 \\
0 & 0   &\cos\theta & 0 \\
0 & 0 & 0   &\cos\theta
\end{array} \right).
\label{eq:sm}
\end{equation}
As an example, consider the case of right--angle
scattering of unpolarized light, $\mathcal{I}' = (I'/2, I'/2, 0, 0)
=(I'/2) (1,1,0,0) $.
Denoting the scattered radiation without a prime,
we have 
\[
\mathcal{I} = \sigma_T {\bf R}\cdot \mathcal{I}' 
%  = \sigma_T I'(\cos^2\theta/2, 1/2, 0, 0) 
%  = \sigma_T I'(0,1/2,0,0),
  = \sigma_T (I'/2) (\cos^2\theta, 1, 0, 0) 
  = \sigma_T (I'/2) (0,1,0,0);
\]
that is, the scattered light is polarized perpendicular to the scattering plane.
We note that in the above expression the dimension of $\mathcal{I}$ is different
from that of $\mathcal{I}'$ (due to the Thomson cross section).
This is because we are considering a single scattering due to one electron.
When one deals with the scatter by an ensemble of particles,
the formal solution of the radiative transfer gives the scattered intensity
with correct dimensionality (see e.g.~\citealt{chandra}, \S 16--17).
{\it For simplicity, however, we shall ignore this inconsistency here and in 
the next section.}

Below it will be important to remember that if there is no $V$ component in
the incident radiation, Rayleigh/Thomson scattering could not induce net
circular polarization in the scattered light.

%%%%%%%%%%%%%%%%%%%%%%%%%%%%%%%%%%%%%%%%%%%%%%%%%%
\subsubsection{Polarization at Last Scattering Surface
\label{sec:lss}}

Suppose the incident radiation field is unpolarized, $Q'=U'=V'=0$.
Without loss of generality, we can choose the $\hat{z}$--axis to be along the
propagation direction of the scattered light, $\hat{k}_s$, and denote the
incident
radiation from $(\theta',\phi')$ by the vector (recall the definition from
last section) $\mathcal{I}'(\theta',\phi') = I'_{\theta',\phi'} (1/2,1/2,0,0)$,
where $I'_{\theta',\phi'}$ is the
total incident intensity. Let us also define the $\hat{x}$-- and $\hat{y}$--axes
to be parallel and perpendicular to the scattering plane. The scattered
intensity vector, defined with respected to this coordinate choice, is then
$\mathcal{I} = (3\sigma_T/16\pi ) I'_{\theta',\phi'} (\cos^2\theta',1,0,0)$. Written
explicitly, $I(\hat{z}) = (3\sigma_T/16\pi ) I'_{\theta',\phi'}
(1+\cos^2\theta')$, $Q(\hat{z}) = (3\sigma_T/16\pi ) I'_{\theta',\phi'}\sin^2\theta'$
and $U(\hat{z}) = 0$.  Note that the notation reminds us that the quantities are
evaluated on the chosen $\hat{z}$--axis.

The above results are for one single scattering. The total scattered intensities
are obtained by integrating over all incident radiation. Since for every
scattering event there is a unique scattering plane, thus coordinates $x-y$
(although $z$ is
always the same), we need to specify a standard coordinate system, referring to
which the integrated $Q$ and $U$ are calculated. In another words,
\begin{eqnarray}
\label{eq:iz}
I(\hat{z}) & = & {3\sigma_T \over 16\pi } \int d\Omega'
    (1+\cos^2\theta') I'_{\theta',\phi'},\\
\label{eq:qz}
Q(\hat{z}) & = & {3\sigma_T \over 16\pi } \int d\Omega'
    \sin^2\theta'\,\cos2\phi'\, I'_{\theta',\phi'},\\
\label{eq:uz}
U(\hat{z}) & = & -{3\sigma_T \over 16\pi } \int d\Omega'
    \sin^2\theta'\,\sin2\phi'\, I'_{\theta',\phi'},
\end{eqnarray}
where Eqn~(\ref{eq:qutransform}) is used, and $-\phi$ is the
azimuthal angle of every scattering plane in the standard coordinates.

If we further expand the incident intensity by spherical harmonics
\[
I'_{\theta',\phi'} = \sum_{l,m} a'_{lm}\,Y_{lm} (\theta',\phi'),
\]
Eqn~(\ref{eq:iz}) becomes
\begin{eqnarray*}
I(\hat{z}) & = & {3\sigma_T \over 16\pi } \sum_{l,m} a'_{lm} \int d\Omega'
    Y_{lm} \left( \sqrt{4\pi} Y_{00} + {1\over3}\left(
    \sqrt{ {16\pi \over 5}} Y_{20} + \sqrt{4\pi} Y_{00}\right) \right)\\
 & = & {\sigma_T \over 4\pi } \left(2\sqrt{\pi} a'_{00} + \sqrt{{\pi \over 5}}
    a'_{20} \right),
\end{eqnarray*}
where we have used the explicit form of $Y_{00}(\theta',\phi') = 1/\sqrt{4\pi}$ and
$Y_{20}(\theta',\phi') = \sqrt{5/16\pi}(3\cos^2\theta'-1)$ in replacing 
the factor ($1 +\cos^2\theta'$) in
the integrand of Eqn~(\ref{eq:iz}). Similarly, because the integrand of $Q-iU$
is directly proportional to $Y_{22}(\theta',\phi') = \sqrt{15/32\pi} 
{\rm e}^{2i\phi'} \sin^2\theta'$, we deduce
\begin{eqnarray*}
(Q-iU)(\hat{z})  =  {3\sigma_T \over 16\pi } \sum_{l,m} a'_{lm} \int d\Omega'
    \sin^2\theta' {\rm e}^{2i\phi'} Y_{lm} (\theta',\phi')
  =   {3\sigma_T \over 4\pi } \sqrt{{2\pi\over 15}}\,a'_{22}.
\end{eqnarray*}
Therefore, if there exist a nonzero quadrupole moment $a'_{22}$ in the
unpolarized incident radiation field, the total scattered radiation in $\hat{z}$
direction would be polarized \citep{k96}. In other words, if we fix a
coordinate system, e.g. celestial coordinates, place ourselves at the origin
and look south, the polarization in the CMB we observe along this direction
was generated by a $a'_{22}$ moment in the incident radiation field of
the electron which last scattered the light we observe.

If we look along any other direction we see light that was scattered not along
the $\hat{z}$ direction by some electron in the last scattering surface but
some other direction $\hat{n}$ which points at us. This $\hat{n}$ makes an angle
$\beta$ with the $\hat{z}$ axis. We can expand the incoming radiation field in
a frame whose $\hat{z}$ axis is pointing toward that direction, and relate the
expansion coefficients with those in the unrotated frame \citep{k99}. 
Denoting the quantities
in this new/rotated frame with a tilde, we have $(\widetilde{Q}-i\widetilde{U})
(\hat{\widetilde{z}}) \propto \widetilde{a'}_{22}$. To relate the coefficients of
multiple expansion in the original/unrotated frame with those in the rotated
frame, we note that
\[
I'(\hat{n}) = \widetilde{I'}(\hat{n}_{rot}) = \sum_{x,y} \widetilde{a'}_{xy}\,Y_{xy} (\hat{n}_{rot}),
\]
where $\hat{n}$ denotes the angle $(\theta'=\beta,\phi')$ in the unrotated
frame, and $\hat{n}_{rot}$ denotes the same direction, but in the rotated
frame; $\hat{n}_{rot} = R(\beta)\,\hat{n}$, where $R$ is the rotation operator.
Then
\begin{eqnarray}
\nonumber
\widetilde{a'}_{xy} & = & \int d\hat{n}_{rot}\,Y^*_{xy}(\hat{n}_{rot}) \widetilde{I'}(\hat{n}_{rot}) \\
\nonumber
 & = & \int d\hat{n}_{rot}\,Y^*_{xy}(R(\beta)\,\hat{n}) I'(\hat{n}) \\
\nonumber
 & = & \int d\hat{n}\,\sum_m \mathcal{D}^x_{ym}(R) Y^*_{xm}(\hat{n}) I'(\hat{n}) \\
\nonumber
 & = & \sum_m \mathcal{D}^x_{ym}(R)\,a'_{xm},
\end{eqnarray}
where $\mathcal{D}^x_{ym}$ is the Wigner D-symbol \citep[e.g.][]{sakurai}. 
Now, because of the 
transformation properties of the spherical harmonics under spatial rotations,
only $a'_{2m}$ components of the 
incidental radiation field would contribute to $\widetilde{a'}_{22}$ which
generates the polarization in the $\hat{n}$ direction. 

As an example
let us consider an azimuthally symmetric incident field, i.e. only $a'_{20} \neq 0$,
then $\mathcal{D}^x_{ym}(R) = d^x_{ym}(\beta)$ and therefore $\widetilde{a'}_{22}
= d^2_{20}(\beta)\,a'_{20} = (\sqrt{6}/4) \,a'_{20} \sin^2 \beta$
\citep{varshalovich}. We obtain \citep{k99}
\[
(Q-iU)(\hat{n}) = {3\sigma_T \over 4\pi } \sqrt{{2\pi\over 15}}\,\widetilde{a'}_{22} 
  = {3\sigma_T \over 8\pi } \sqrt{{\pi\over 5}}\,a'_{20}\,\sin^2\beta.
\]
This means that the polarization due to an $a_{20}$ component in the incident
radiation field is maximal if we look along the equator, since
$\beta=\pi/2$. In that case we are seeing the incident radiation field ``edge on''.
Notice also, that because the incident field is real, $a'_{20}$ is real, and hence $U(\hat{n})=0$.
In this way the contribution from different $a_{2m}$ modes of the incident
radiation field to the $Q$ and $U$ we observe varies depending on our line of sight.

\begin{figure}
%\figurenum{A2}
\epsscale{1.}
%\plottwo{y10.ps}{y20.ps}
\plottwo{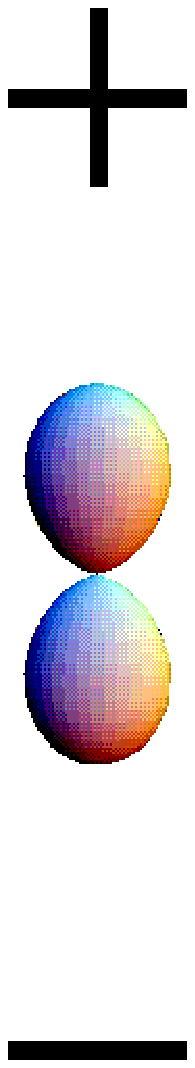}{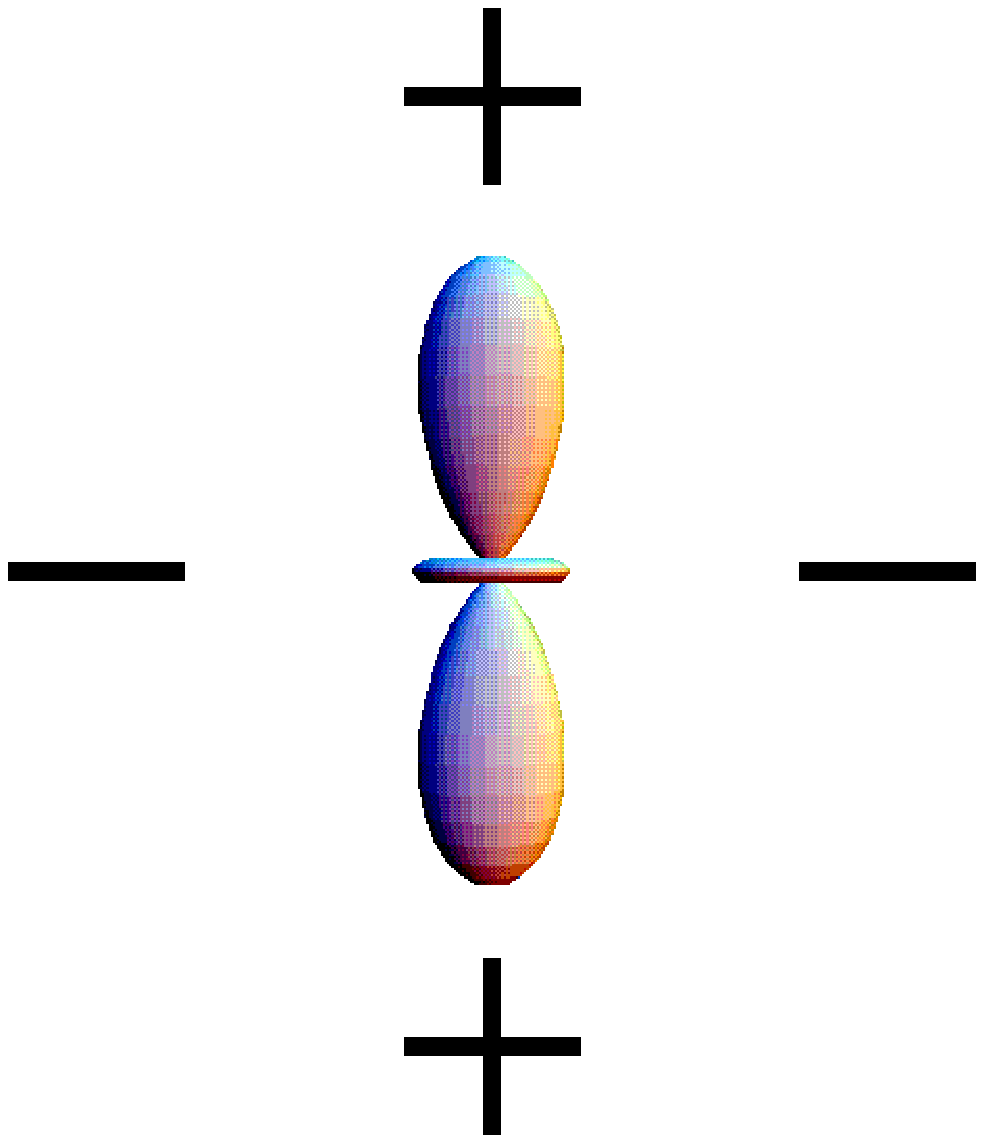}
\caption{
        By symmetry, a uniform (monopole) radiation field could not produce any
        polarization.  {\it Left}: A dipole radiation field. The direction of
	the out-going radiation is out of the paper. The plus and minus
        signs indicate the
        intensity of the radiation. The light coming from both top and bottom
        will be polarized perpendicularly to the scattering plane, i.e. along 
	the horizontal direction. The light from the sides will be polarized
        vertically. Since the sum of the incident light from  vertical 
	directions is equal to the sum of the incident light from horizontal 
	directions, the net polarization is zero. {\it Right}: A quadrupole
        radiation field. The direction of the out-going radiation is also
	out of the paper. The intensity from both the top and the
        bottom is higher than that from the sides. Therefore the outgoing
        light has a net polarization
        in the  horizontal direction.
    }
\label{fig:mondip}
\end{figure}

There is a pictorial way to infer why the existence of polarization 
requires a quadrupole moment in the incident radiation field. Referring to
Fig~\ref{fig:mondip}  we see that for isotropic incident radiation (monopole),
there is no polarization, by symmetry.  For dipole anisotropy the amplitude
from the top is 
greatest, while that from bottom is weakest, and average on the sides. The polarization directions 
are horizontal for scattering from the
top and bottom and vertical from the sides. The radiation from the top would
produce horizontally polarized light, but the lack of radiation from the  bottom
 cancels this effect. The net outgoing polarization is zero. We need to
go to quadrupole moments to obtain polarization. In the configuration shown in
the figure, we have a net outgoing horizontal polarization, perpendicular to the long lobes.

Long before recombination,  in thermal equilibrium, 
both polarization states of photons are equally populated; therefore the incident intensity
field should not possess any polarization. In particular, $V=0$. Since Thomson
scattering can not generate circular polarization, as discussed in previous
section, only $Q$ and $U$ polarization is expected to be present in the CMB.

In this section we assumed that the incident radiation was unpolarized, which
is true before recombination. In this case polarization is generated
by a quadrupole moment in the incident radiation intensity. If we do
allow polarization for the last few scatters before the photons begin to
free-stream, there are two additional ways polarized emission can occur:
polarized monopole and a polarized quadrupole. These two latter effects are
subdominant to the first and we will not describe them in detail
here. However, we do keep the polarization monopole and quadrupole terms when
we discuss the solution of the cosmological photon Boltzmann equation in
section \ref{sec:scalarpert}. 

Gravitational wave perturbations can produce
outgoing polarized radiation in additional ways which leads to additional
terms coupling to the $a_{00}$ component of the incident field as well as
higher order terms. We include in our calculation of tensor perturbations in \ref{sec:tensorpert}. 

%%%%%%%%%%%%%%%%%%%%%%%%%%%%%%%%%%%%%%%%%%%%%%%%%%
%%%%%%%%%%%%%%%%%%%%%%%%%%%%%%%%%%%%%%%%%%%%%%%%%%

\section{Temperature and Polarization Power Spectra}

Having outlined how polarization is generated at the last scattering surface
we now turn to the prediction of the statistical properties of the polarized
CMB. Since the primordial perturbations are expected to be Gaussian to a high
degree of accuracy and since linear theory is a highly accurate approximation
to the evolution of these perturbations until last scattering, the
small anisotropies of temperature and polarization in the CMB are expected to
follow Gaussian statistics. That smallness of the
anisotropies is also good evidence for global isotropy and homogeneity of the
cosmos. These two aspects, Gaussianity and isotropy lead us to believe that
the statistics of the anisotropies are 
comprehensively summarized in the \emph{power spectra} of the
perturbations. These are therefore the central objects of investigation in all
current missions observing CMB temperature and polarization.

In this section we follow closely the work of ZS97 for the calculations of the
temperature and polarization power spectra of the CMB.\footnote{Some typographical errors in ZS97 have been corrected.}
With the use of the
spin--weighted spherical harmonics, ZS97 unified the treatments of these 
spectra. We first introduce the basic quantities necessary for the calculations
of the perturbations that leave their signature on the CMB (\S~\ref{sec:defin}),
then detail the calculations of the spectra due to the scalar perturbations
(\S~\ref{sec:scalarpert}) and tensor perturbations (\S~\ref{sec:tensorpert}).

%%%%%%%%%%%%%%%%%%%%%%%%%%%%%%%%%%%%%%%%%%%%%%%%%%
\subsection{Definitions
\label{sec:defin}}

Suppose we are interested in the temperature and polarization anisotropy
of CMB at a point ($\theta,\phi$) on the sky. The unit vector
along the line of sight is called $\hat{n}$. The Stokes parameters are measured with respect to the local coordinate system specified by the vectors $(\hat{{\rm e}}_\theta,\hat{{\rm e}}_\phi)$ on the tangent plane at the point of interest.
From Eqn~(\ref{eq:qutransform}) we see
that, under a rotation through an angle $\psi$ about $\hat{n}$, a
particular combination of Stokes $Q(\hat{n})$ and $U(\hat{n})$ transforms as
\begin{equation}
(Q\pm iU)'(\hat{n}) = {\rm e}^{\mp2i\psi} (Q\pm iU)(\hat{n}),
\end{equation}
i.e., like spin--$\pm2$ quantities. We therefore can expand them by the
spin--2 spherical harmonics:
\begin{eqnarray}
\nonumber
(Q+ iU)(\hat{n}) & = & \sum_{l,m}\,a_{2,lm}\,_2Y_{lm}(\hat{n}), \\
(Q- iU)(\hat{n}) & = & \sum_{l,m}\,a_{-2,lm}\,_{-2}Y_{lm}(\hat{n}).
\end{eqnarray}
On the other hand, the temperature anisotropy is expanded by the usual 
spherical harmonics
\begin{eqnarray}
T(\hat{n}) & = & \sum_{l,m}\,a_{T,lm}\,Y_{lm}(\hat{n}),
\end{eqnarray}
since it is invariant under rotation (spin--0).

With the help of the spin raising and lowering operators, together with 
Eqn~(\ref{eq:additional}), we obtain two spin--$0$ quantities:
\begin{eqnarray}
\label{eq:qiu_ylm}
\nonumber
\flat^2\,(Q+ iU)(\hat{n}) & = & \sum_{l,m}\,a_{2,lm}\,\flat\flat\,_2Y_{lm}(\hat{n})\\
\nonumber
 &  = & \sum_{l,m}\,a_{2,lm}\,\flat \left( -\sqrt{(l+2)(l-1)}\,_1Y_{lm} \right) \\
\nonumber
 & = & \sum_{l,m}\,\left[ {(l+2)!\over(l-2)!} \right]^{1/2}
    a_{2,lm}\,Y_{lm}(\hat{n}), \\
\sharp^2\,(Q- iU)(\hat{n}) & = & \sum_{l,m}\,\left[ {(l+2)!\over(l-2)!} \right]^{1/2}
    a_{-2,lm}\,Y_{lm}(\hat{n}).
\end{eqnarray}

From these the expansion coefficients can be found by using the orthogonality of spin--$0$ \& $2$ spherical harmonics:
%Using the orthogonalities of spin--$0$ \& $2$ spherical harmonics and Eqn~(\ref{eq:qiu_ylm}), we obtain the expansion coefficients:
\begin{eqnarray}
\nonumber
a_{T,lm} & = & \int\,d\Omega\,Y_{lm}^*(\hat{n})\,T(\hat{n}), \\
\nonumber
a_{2,lm} & = & \int\,d\Omega\,_2Y_{lm}^*(\hat{n})\,(Q+ iU)(\hat{n}), \\
\nonumber
 & = & \left[ {(l-2)!\over(l+2)!} \right]^{1/2} \int\,d\Omega\,Y_{lm}^*(\hat{n})\,
    \flat^2\,(Q+ iU)(\hat{n}), \\
\nonumber
a_{-2,lm} & = & \int\,d\Omega\,_{-2}Y_{lm}^*(\hat{n})\,(Q- iU)(\hat{n}), \\
 & = & \left[ {(l-2)!\over(l+2)!} \right]^{1/2} \int\,d\Omega\,Y_{lm}^*(\hat{n})\,
    \sharp^2\,(Q- iU)(\hat{n}).
\label{eq:as}
\end{eqnarray}

In Appendix \ref{sec:parity} we discuss the parity of $Q$ and $U$ under spatial
inversion, the operation  that takes $\hat{n}$ into $-\hat{n}$. We find that 
$Q$ is parity even, $U$ is parity odd.
In the following we group together quantities of the same parity, and
work with two spin--0 quantities $\widetilde{E}(\hat{n})$ and
$\widetilde{B}(\hat{n})$ constructed from $\flat^2\,(Q+ iU)(\hat{n})$ and
$\sharp^2\,(Q- iU)(\hat{n})$ (ZS97):
\begin{eqnarray}
\nonumber
\widetilde{E}(\hat{n}) & \equiv & {-1\over2} \left( \flat^2\,(Q+ iU)(\hat{n})
    + \sharp^2\,(Q- iU)(\hat{n}) \right) \\
\nonumber
 & = & \sum_{l,m} \left[ {(l+2)!\over(l-2)!} \right]^{1/2}\,a_{E,lm}\,Y_{lm}(\hat{n}) \\
\label{eq:ewidetilde}
 & \equiv & \sum_{l,m} \,a_{\widetilde{E},lm}\,Y_{lm}(\hat{n}), \\
\nonumber
\widetilde{B}(\hat{n}) & \equiv & {-1\over2i} \left( \flat^2\,(Q+ iU)(\hat{n})
    - \sharp^2\,(Q- iU)(\hat{n}) \right) \\
\nonumber
 & = & \sum_{l,m} \left[ {(l+2)!\over(l-2)!} \right]^{1/2}\,a_{B,lm}\,Y_{lm}(\hat{n}) \\
\label{eq:bwidetilde}
 & \equiv &  \sum_{l,m} \,a_{\widetilde{B},lm}\,Y_{lm}(\hat{n}),
\end{eqnarray}
where we have introduced coefficients which are linear combinations of
$a_{2,lm}$ and $a_{-2,lm}$:
\begin{eqnarray}
\nonumber
a_{E,lm} & \equiv & -(a_{2,lm}+a_{-2,lm})/2, \\
\nonumber
a_{B,lm} & \equiv  & -(a_{2,lm}-a_{-2,lm})/2i, \\
a_{(\widetilde{E},\widetilde{B}),lm} & = & \left[ {(l+2)!\over(l-2)!} \right]^{1/2} a_{(
E,B),lm}.
\end{eqnarray}

 The main advantage of working with $\widetilde{E}$ and $\widetilde{B}$ compared
to $Q$ and $U$ is that they
are scalar, spin-$0$ quantities. In this way the
calculation of polarization spectra proceeds analogously to that of the temperature spectrum. While
the same is true for the quantities defined in Eqn~(\ref{eq:qiu_ylm}), other
advantages of $\widetilde{E}$ and $\widetilde{B}$ are that they are real, and,
more importantly, that they have  
distinct parities (see Appendix~\ref{sec:parity}). Under parity transformation
the $\widetilde{E}$ pattern remains the same, while that of $\widetilde{B}$ changes
sign. It is this property that reduces the number of power spectra (see below)
that we need to calculate: cross--correlations between $\widetilde{B}$ and $T$ or
$\widetilde{E}$ vanish \footnote{Equivalently, one can define
$E(\hat{n})  =  \sum \,a_{E,lm}\,Y_{lm}(\hat{n})$ and
$B(\hat{n})  =  \sum \,a_{B,lm}\,Y_{lm}(\hat{n})$,
but these are almost identical to $\widetilde{E}$ and $\widetilde{B}$ defined above.}
(see \S~\ref{sec:parity}).

Finally, let us define the power spectra necessary to describe the statistics of
CMB temperature and polarization maps:
\begin{eqnarray}
%\nonumber
%C_{Tl} & = & {1\over 2l+1} \sum_m \langle a_{T,lm}^*\,a_{T,lm} \rangle, \\
%\nonumber
%C_{El} & = & {1\over 2l+1} \sum_m \langle a_{E,lm}^*\,a_{E,lm} \rangle, \\
\nonumber
C_{Xl} & = & {1\over 2l+1} \sum_m \langle a_{X,lm}^*\,a_{X,lm} \rangle, \\
C_{Cl} & = & {1\over 2l+1} \sum_m \langle a_{T,lm}^*\,a_{E,lm} \rangle,
\end{eqnarray}
where $X\,=\,T,\,E,\,B$, and $\langle ... \rangle$ denotes ensemble average.
From Eqn~(\ref{eq:as}) we see that
\[
\langle a_{T,l'm'}^*\,a_{T,lm} \rangle = \int \int d\Omega' d\Omega
  \,\langle T^*(\hat{n}') T(\hat{n})\rangle\,Y_{l'm'}(\hat{n}') Y_{lm}(\hat{n}).
\]
If correlation in temperature at different sky positions $\hat{n}'$ and
$\hat{n}$ can be expressed as $\langle T^*(\hat{n}') T(\hat{n})\rangle = 
C_{TT}(\hat{n}' \cdot \hat{n})$, i.e. only a function of the angle between the
two position vectors, we can expand the correlation function in terms of the
Legendre polynomial: 
\[
C_{TT}(\hat{n}' \cdot \hat{n}) = \sum_q \frac{2q+1}{4\pi} C_{Tq} P_q(\hat{n}' \cdot \hat{n}).
\]
With the help of the addition theorem of spherical harmonics 
\[
P_q(\hat{n}' \cdot \hat{n}) = \frac{4\pi}{2q+1} \sum_k Y_{qk}^*(\hat{n}') Y_{qk}(\hat{n}),
\]
we obtain
\begin{eqnarray}
\nonumber
\langle a_{T,l'm'}^*\,a_{T,lm} \rangle & = & \sum_{q,k} C_{Tq} \int d\Omega' \,
  Y_{qk}^*(\hat{n}') Y_{l'm'}(\hat{n}') \int d\Omega \,Y_{qk}(\hat{n}) 
  Y_{lm}(\hat{n}) \\
\nonumber
  & = & \sum_{q,k} C_{Tq}\,\delta_{ql'} \delta_{km'} \delta_{ql} \delta_{km} \\
  & = & C_{Tl}\, \delta_{ll'} \delta_{mm'}.
\end{eqnarray}
Similar calculations lead to other correlations between $T$, $E$, and $B$ modes;
we therefore conclude
\begin{eqnarray}
\nonumber
\langle a_{X,l'm'}^*\,a_{X,lm} \rangle & = & C_{Xl}\,\delta_{l'l}\,\delta_{m'm}, \\
\nonumber
\langle a_{T,l'm'}^*\,a_{E,lm} \rangle & = & C_{Cl}\,\delta_{l'l}\,\delta_{m'm}, \\
\langle a_{B,l'm'}^*\,a_{T,lm} \rangle & = &\langle a_{B,l'm'}^*\,a_{E,lm} \rangle = \ 0.
\end{eqnarray}

%%%%%%%%%%%%%%%%%%%%%%%%%%%%%%%%%%%%%%%%%%%%%%%%%%
\subsection{Scalar Perturbations}
\label{sec:scalarpert}

Scalar perturbations (e.g.~perturbations of the gravitational potential, see
\citealt{efstathiou90}, \citealt{ma95}, and \citealt{dodelson} for detailed treatments and \citealt{bertschinger00} for a useful summary of structure formation)  can be expanded into plane waves, each
characterized by a  
wave vector ${\bf k}$. We can decompose the  temperature anisotropy seen by an
observer at conformal time $\tau$ into individual contributions from different
$\mathbf{k}$-modes of the scalar perturbation. There are only two relevant
directions,  $\hat{k}$ and the line of sight $\hat{n}$. Without loss of
generality we can always choose coordinates such that  $\hat{z} \parallel {\bf
  k}$. We denote this temperature anisotropy by 
$\Delta_T^s(\tau, k, \mu)$,
which by the above argument only depends on the angle between the photon direction $\hat{n}$ and
$\hat{k}$, $\mu \equiv \hat{n} \cdot \hat{k}$, the amplitude of the mode,
$k \equiv |{\bf k}|$, and the conformal time $\tau$.
The use of ``$\Delta$''  means the anisotropy is due
to this single mode; and the superscript ``s'' denotes the scalar perturbation.

Unlike the case of temperature anisotropy,
for the polarization calculations, we need to specify a local coordinate system
for every point on the sky.
A ``natural'' choice is the local unit vectors $\hat{e}_\theta$ and
$\hat{e}_\phi$ of the usual spherical coordinates, together with $\hat{n}$ 
as the third unit vector, as mentioned in \S~\ref{sec:defin}. This
is a good choice because of two symmetries in the problem: in addition to the
azimuthal symmetry, there is also reflection symmetries, that is, space inverse
with respect to the plane containing ${\bf k}$ and $\hat{n}$. Under space
inverse, $Q$ remains unchanged while $U$ changes sign (see \S~\ref{sec:parity}).
%also Fig~2.5 of \citealt{z98}).
Therefore, under this choice of coordinates, the polarization can only be in
either $\hat{e}_\theta$ or $\hat{e}_\phi$ direction, that is, purely $Q$ with
$U=0$. Since $V$ is also zero,
the polarization tensor is diagonal in this frame (c.f. Eqn~\ref{eq:phomatrix}).

In short, simple symmetry considerations suggest that only the $Q$ component
will be present (which we denote as $\Delta_Q^s$), and its
amplitude depends only on $\mu$, $k$ and $\tau$.
We denote the polarization due to this
single mode by $\Delta_P^s(\tau, k, \mu) \equiv \Delta_Q^s+i\Delta_U^s$. % = Q+i U$.
Specifically, $\Delta_P^s = \Delta_Q^s$ in this frame.

The evolution of the anisotropies is described by the Boltzmann equation.
In the synchronous gauge, we have (e.g., \citealt{ma95})
\begin{eqnarray}
\nonumber
\dot{\Delta}_T^s + ik\mu \Delta_T^s & = & -{1\over6}\dot{h}
  -{1\over3}(\dot{h}+6\dot{\eta}) P_2(\mu) 
%\nonumber
% & & 
%  +\dot{\kappa} \left[ -\Delta_T^s + \Delta_{T0}^s + i\mu v_b
  +\dot{\kappa} \left[ -\Delta_T^s + \Delta_{T0}^s + \hat{n} \cdot {\bf v_b}
     -{1\over2} P_2(\mu) \Pi \right], \\
\nonumber
\dot{\Delta}_P^s + ik\mu \Delta_P^s & = & \dot{\kappa} \left[ -\Delta_P^s
    + {3\over4}(1-\mu^2) \Pi \right], \\
\Pi(\tau,k) & = & \Delta_{T2}^s +\Delta_{P0}^s +\Delta_{P2}^s,
\label{eq:boltz}
\end{eqnarray}
where $h$ and $h+6\eta$ are the trace and traceless scalar parts of the metric 
perturbation, ${\bf v_b}$ is the baryon velocity, and the
$\Delta^s_{(T,P)j}$ ($j=0,2$) are from the Legendre expansion
\[
\Delta(k,\mu) = \sum_l (-i)^l (2l+1) \Delta_l(k) P_l(\mu).
\]
%($P_l(\mu)$ is the Legendre polynomial of order $l$; in particular,
%$P_2(\mu) = (3\mu^2-1)/2$). 
Using this expansion convention, $\hat{n} \cdot {\bf v_b} = -i v_b \mu$.
The differential optical depth is defined through
\begin{equation}
\dot{\kappa} = a(\tau)\,n_e\,x_e\,\sigma_T, \ \ \ \ \ \
  \kappa(\tau) = \int_\tau^{\tau_0} \dot{\kappa}(\tau')\,d \tau',
\label{eq:kappa}
\end{equation}
where $a(\tau)$ is the cosmic expansion factor, $n_e$ is the electron density
and $x_e$ is the ionization fraction.
Note the time derivatives are all with respect to $\tau$.

We caution that $\Delta_T^s$ defined here is the perturbation of
photon brightness temperature, which is $1/4$ of the perturbation in
intensity; the latter is adopted by \citet{be84} with the same notation.
The equation (1) in \citet{be84} is therefore 4 times Eqn
(\ref{eq:boltz}). Also notice $h_{33}$ in \citet{be84} is $h+4\eta$ used
here.

Note also that the three contributions to $\Pi$, the term which sources 
polarized photon brightness in Eqn~(\ref{eq:boltz}), correspond to the 
contributions discussed in section \ref{sec:lss}.

In Appendix~\ref{sec:soln} we show how to write down an integral solution to the Boltzmann equation \citep{sz96}.  This method of solving the Boltzmann equation is called the line-of-sight technique, as a way of distinguishing it from traditional techniques such as the moment expansion. The result is
\begin{eqnarray}
\nonumber
\Delta_T^s(\tau_0, k, \mu) & = & \int_0^{\tau_0} d\tau\,{\rm e}^{-ix\mu}
    S_T^s(\tau,k), \\
\nonumber
S_T^s(\tau,k) & = & {\rm e}^{-\kappa} (\dot{\eta} +\ddot{\alpha}) + 
  g(\tau) \left( \Delta_{T0}^s +2\dot{\alpha} +
  {\dot{v_b}\over k} +{\Pi\over4} + {3\ddot{\Pi} \over 4k^2} \right) \\
 & &  + \dot{g}(\tau) \left( \alpha + {v_b \over k} + {3\dot{\Pi} \over 2k^2} \right)
\nonumber
  +{3\Pi \ddot{g}(\tau) \over 4 k^2}, \\
\nonumber
g(\tau) & = & \dot{\kappa}\,{\rm e}^{-\kappa}, \\
\Delta_P^s(\tau_0, k, \mu) & = & {3\over4}(1-\mu^2)
    \int_0^{\tau_0} d\tau\,{\rm e}^{-ix\mu} \,g(\tau)\,\Pi(\tau,k), 
\label{eq:bsoln}
\end{eqnarray}
where the auxiliary variable $x = k(\tau_0-\tau)$, and 
$\alpha = (\dot{h}+6\dot{\eta})/2k^2$. In particular, the visibility function
$g(\tau)$ encodes the ionization/recombination history of the Universe and is
strongly peaked at the epoch of recombination. Also notice how simple the $\mu$
dependences are in these solutions.

So far all the calculations are done for one ${\bf k}$ mode; to obtain the total
contribution from all modes we need to sum over their contributions. Consider
the temperature anisotropy first. At the direction $\hat{n}$ the contribution
from scalar perturbations is
\begin{equation}
T^s(\hat{n}) = \int\,d^3{\bf k}\,\xi({\bf k})\,\Delta_T^s(\tau_0, k, \mu),
\label{eq:temp}
\end{equation}
where $\xi({\bf k})$ is a random variable describing the stochastic property of
the initial density field; it satisfies
\begin{equation}
\langle \xi^*({\bf k}_1)\,\xi({\bf k}_2) \rangle = P^s(k)\,\delta({\bf k}_1-{\bf k}_2),
\end{equation}
where $P^s(k)$ is the primordial power spectrum for scalar perturbations.
From this we can calculate the power spectrum $C_{Tl}$ (ZS97):
\begin{eqnarray}
\nonumber
C_{Tl} & = & {1\over 2l+1} \sum_m \langle a_{T,lm}^*\,a_{T,lm} \rangle \\
\nonumber
 & = & {1\over 2l+1} \sum_m \langle
  \left( \int\,d\Omega'\,Y_{lm}^*(\hat{n}')\,T^s(\hat{n}') \right)^*
  \left( \int\,d\Omega\,Y_{lm}^*(\hat{n})\,T^s(\hat{n}) \right) \rangle \\
\nonumber
 & = & {1\over 2l+1} \sum_m \int\,d\Omega'\,d\Omega\,d^3{\bf k}'\,d^3{\bf k}
  \,Y_{lm}(\hat{n}')\,Y_{lm}^*(\hat{n})\, \\
\nonumber
 & & \mbox{\ \ \ \ \ \ \ \ \ \ \ \ \ \ \  }
  \times \Delta^{s*}_T(\tau_0,k',\mu')\,\Delta^s_T(\tau_0,k,\mu)\,
  \langle \xi^*({\bf k}')\,\xi({\bf k}) \rangle \\
\nonumber
 & = & {1\over 2l+1} \sum_m \int d^3{\bf k}\,P^s(k) \int\,d\Omega'\,d\Omega\,
  Y_{lm}(\hat{n}')\,Y_{lm}^*(\hat{n}) \\
\nonumber
 & & \mbox{\ \ \ \ \ \ \ \ \ \ \ \ \ \ \  }
  \times \Delta^{s*}_T(\tau_0,k,\hat{n}'\cdot \hat{k})
  \,\Delta^s_T(\tau_0,k,\hat{n}\cdot \hat{k}) \\
\nonumber
 & = & {1\over 2l+1} \int d^3{\bf k}\,P^s(k) \sum_m
  \left| \int\,d\Omega\,Y_{lm}^*(\hat{n}) \Delta^s_T(\tau_0,k,\mu) \right|^2 \\
%\nonumber
% & = & {1\over 2l+1} \int d^3{\bf k}\,P^s(k) \sum_m
%  \left| \int\,d\Omega\,Y_{lm}^*(\hat{n}) \int_0^{\tau_0} d\tau\,
%  {\rm e}^{-ix\mu} S_T^s(\tau,k)\, \right|^2 \\
\nonumber
% & = & {1\over 2l+1} \int d^3{\bf k}\,P^s(k) \sum_m
%  | \int_0^{\tau_0} d\tau \, \underbrace{\int\,d\Omega\, Y_{lm}^*(\hat{n})
%  \,{\rm e}^{ix\mu}}_{\sqrt{4\pi(2l+1)}i^l\,j_l(x)\,\delta_{m0}} % \footnote{blah}}
%  \,S_T^s(\tau,k) |^2 \\
 & = & {1\over 2l+1} \int d^3{\bf k}\,P^s(k) \sum_m
  \left| \int_0^{\tau_0} d\tau \,S_T^s(\tau,k)\,\int d\Omega\, Y_{lm}^*(\hat{n})
  \,{\rm e}^{-ix\mu} \,\right|^2 \\
\nonumber
 & = & (4\pi)^2 \int k^2\,dk\,P^s(k) \left[ \int_0^{\tau_0}
  d\tau \,S_T^s(\tau,k)\,j_l(x) \right]^2 \\
 & = & (4\pi)^2 \int k^2\,dk\,P^s(k) [\Delta^s_{Tl}(k)]^2,
\label{eq:ctlscalar}
\end{eqnarray}
where
\begin{equation}
\Delta^s_{Tl}(k) \equiv \int_0^{\tau_0}d\tau \,S_T^s(\tau,k)\,j_l(x),
\label{eq:ctlscalarsource}
\end{equation}
and $j_l(x)$ is the spherical Bessel function of order $l$. 
Because of the configuration that $\hat{z} \parallel {\bf k}$,
the identity $\int d\Omega\,Y_{lm}^*(\hat{n})\,{\rm e}^{-ix\mu} = 
[4\pi(2l+1)]^{1/2}\,(-i)^l\, j_l(x)\,\delta_{m0}$ has been used.

Next let us consider the case for polarization. Because under a rotation of the
plane perpendicular to $\hat{n}$, the Stokes $Q$ and $U$ change into each other,
we need to refer all calculations to a fixed coordinate. In other words, the
contributions to polarization anisotropy from all scalar modes are
\[
%(\Delta_Q^s\pm i\Delta_U^s)^s(\hat{n}) = \int\,d^3{\bf k}\, \xi({\bf k})\,
(Q\pm i U)^s(\hat{n}) = \int\,d^3{\bf k}\, \xi({\bf k})\,
        {\rm exp}(\mp 2i\phi_{\hat{n},\hat{k}})\, \Delta^s_P(\tau_0,k,\mu),
\]
where $\phi_{\hat{n},\hat{k}}$ is the angle needed to rotate each
$\hat{z} \parallel {\bf k}$ frame to the standard reference frame.
This is a main source of complications associated with polarization spectra
calculations (ZS97). Instead, in complete analogy to the temperature 
calculations (Eqn~\ref{eq:temp}), we use (ZS97)
\begin{equation}
\widetilde{E}^s(\hat{n}) = \int\,d^3{\bf k}\,\xi({\bf k})\,
    \Delta^s_{\widetilde{E}}(\tau_0,k,\mu)
\end{equation}
to compute the power spectra, where the spin--$0$ quantity $\Delta^s_{\widetilde{E}}$
is defined by Eqn~(\ref{eq:ewidetilde}):
\begin{eqnarray}
\nonumber
\Delta^s_{\widetilde{E}}(\tau_0,k,\mu) & = & {-1\over2}
        \left( \flat^2 \Delta^s_P(\tau_0,k,\mu) +
    \sharp^2 \Delta^{s*}_P(\tau_0,k,\mu) \right) \\
\nonumber
 &  = & {-3\over4} \int_0^{\tau_0} d\tau\,g(\tau)\,\Pi(\tau,k)
        \partial_\mu^2 \left[ (1-\mu^2)^2\,{\rm e}^{-ix\mu}\right] \\
 & = & {3\over4} \int_0^{\tau_0} d\tau\,g(\tau)\,\Pi(\tau,k)
        (1+\partial_x^2)^2 \left( x^2\,{\rm e}^{-ix\mu}\right),
\end{eqnarray}
%where we have used the fact that $\Delta^s_P(\tau_0,k,\mu) = \Delta_Q^s+i\Delta_U^s =
%\Delta^{s*}_P(\tau_0,k,\mu) = \Delta_Q^s-i\Delta_U^s = \Delta_Q^s$ and applied
where we have used the fact that $\Delta^s_P(\tau_0,k,\mu) = 
\Delta^{s*}_P(\tau_0,k,\mu)$ (recall that $\Delta_U^s = 0$) and applied
Eqns~(\ref{eq:f2}) \& (\ref{eq:s2}) to the integral solution Eqn~(\ref{eq:bsoln}).
In obtaining the last expression, we have used the fact that
$g(\partial_b)\,[f(b)\,{\rm exp}(-iab)] = g(\partial_b)\,[f(i \partial_a)
\,{\rm exp}(-iab)] = f(i \partial_a)\,[g(\partial_b)\,\,{\rm exp}(-iab)]$ (where
$f(y)$ and $g(y)$ are polynomials of $y$) to simplify the $\mu$ dependence.
It is apparent from the definition Eqn~(\ref{eq:bwidetilde}) that
\[
\Delta^s_{\widetilde{B}}(\tau_0,k,\mu) = {-1\over2i}
        \left( \flat^2 \Delta^s_P(\tau_0,k,\mu)
    - \sharp^2 \Delta^{s*}_P(\tau_0,k,\mu) \right) = 0,
\]
thus
\[
\widetilde{B}^s(\hat{n}) = \int\,d^3{\bf k}\,\xi({\bf k})\,
    \Delta^s_{\widetilde{B}}(\tau_0,k,\mu) = 0.
\]
Therefore the only polarization related power spectra for scalar perturbations
that we need to calculate are $C_{El}$ and $C_{Cl}$.

The calculation of $C_{El}$ is now very similar to that of $C_{Tl}$:
\begin{eqnarray*}
C_{El} & = & {1\over 2l+1} \sum_m \langle a_{E,lm}^*\,a_{E,lm} \rangle \\
 & = & {1\over 2l+1} {(l-2)!\over(l+2)!}
% & &  \mbox{\ \ } \times \sum_m \langle
  \times \sum_m \langle
  \left( \int\,d\Omega'\,Y_{lm}^*(\hat{n}')\,\widetilde{E}(\hat{n}') \right)^*
  \left( \int\,d\Omega\,Y_{lm}^*(\hat{n})\,\widetilde{E}(\hat{n}) \right) \rangle \\
 & = & {1\over 2l+1} {(l-2)!\over(l+2)!} \sum_m
  \int\,d\Omega'\,d\Omega\,d^3{\bf k}'\,d^3{\bf k}
  \,Y_{lm}(\hat{n}')\,Y_{lm}^*(\hat{n})\, \\
 & & \mbox{\ \ \ \ \ \ \ \ \ \ \ \ \ \ } \times
  \Delta^{s*}_{\widetilde{E}}(\tau_0,k',\mu')\,\Delta^s_{\widetilde{E}}(\tau_0,k,\mu)\,
  \langle \xi^*({\bf k}')\,\xi({\bf k}) \rangle \\
 & = & {1\over 2l+1} {(l-2)!\over(l+2)!} \sum_m \int d^3{\bf k}\,P^s(k)
  \int\,d\Omega'\,d\Omega\,Y_{lm}(\hat{n}')\,Y_{lm}^*(\hat{n})
% & & \mbox{\ \ \ \ \ \ \ \ \ \ \ \ \ \ }
% \times
  \Delta^{s*}_{\widetilde{E}}(\tau_0,k,\hat{n}'\cdot \hat{k})
  \,\Delta^s_{\widetilde{E}}(\tau_0,k,\hat{n}\cdot \hat{k}) \\
 & = & {1\over 2l+1} {(l-2)!\over(l+2)!} \int d^3{\bf k}\,P^s(k)
% & & \mbox{\ \ \ \ \ \ \ \ \ \ \ \ \ \ } \times \sum_m \left| {3\over4}
  \sum_m \left| {3\over4}
  \int\,d\Omega Y_{lm}^*(\hat{n})  \int_0^{\tau_0} d\tau\,g(\tau)\,\Pi(\tau,k)
  (1+\partial_x^2)^2 \left( x^2\,{\rm e}^{-ix\mu}\right) \right|^2 \\
 & = & (4\pi)^2 {(l-2)!\over(l+2)!} \int k^2 dk P^s(k)
% & & \mbox{\ \ } \times\,
  \left( {3\over4} \int_0^{\tau_0} d\tau \, g(\tau)\,\Pi(\tau,k)
  ( [1+\partial_x^2]^2[x^2\,j_l(x)]) \right)^2,
\end{eqnarray*}
where we have interchanged the order between angular and temporal integrations,
and performed the angular integration to obtain the spherical Bessel function. To proceed,
we use the differential equation that defines $j_l(x)$ 
\[ j_l''+2j_l'/x +\left( 1-l(l+1)/x^2 \right) j_l = 0, \]
(where the primes denote the derivatives with respect to $x$) to 
replace the $j_l''$ and $j_l'$ terms by $j_l$ (ZS97):
\begin{eqnarray}
\nonumber
(1+\partial_x^2)^2[x^2\,j_l(x)] & = & (1+\partial_x^2) (x^2\,j_l+\partial_x (2x j_l+x^2 j_l')) \\
\nonumber
 & = & (1+\partial_x^2) (2j_l+2x j_l'+l(l+1) j_l) \\
\nonumber
 & = & (2+l(l+1))j_l + 2x j_l' + \partial_x ((4+l(l+1))j_l'-4j_l'-2(x-l(l+1)/x)j_l) \\
\nonumber
 & = & l(l+1)(l(l+1)-2)j_l/x^2 \\
 & = & (l-1)l(l+1)(l+2) j_l/x^2.
\label{eq:j}
\end{eqnarray}
The power spectrum then becomes (ZS97)
\begin{eqnarray}
\nonumber
C_{El} & = & (4\pi)^2 {(l+2)!\over(l-2)!} \int k^2\,dk\,P^s(k) \left[ {3\over4}
\int_0^{\tau_0} d\tau g(\tau)\,\Pi(\tau,k) \,{j_l(x)\over x^2} \right]^2 \\
 & = & (4\pi)^2 \int k^2\,dk\,P^s(k) [\Delta^s_{El}(k)]^2,
\end{eqnarray}
        where
\begin{eqnarray}
\nonumber
\Delta^s_{El}(k) & \equiv  &\left( {(l+2)!\over (l-2)!} \right)^{1/2}
  \int_0^{\tau_0}d\tau \,S_E^s(\tau,k)\,j_l(x), \\
S_E^s(\tau,k)  & \equiv & {3\over4}{g(\tau)\Pi(\tau,k) \over x^2},
\end{eqnarray}

Following similar procedures we obtain the cross correlation power spectrum
\begin{equation}
C_{Cl} = (4\pi)^2 \int k^2 dk\,P^s(k)\,\Delta^s_{Tl}\,\Delta^s_{El}.
\end{equation}

%%%%%%%%%%%%%%%%%%%%%%%%%%%%%%%%%%%%%%%%%%%%%%%%%%
\subsection{Tensor Perturbations}
\label{sec:tensorpert}

The calculations of power spectra due to tensor perturbations (gravity waves)
are basically the same, if somewhat more complicated. The reason is that, for
each Fourier mode of the gravity waves, there are two independent
``polarizations'', often denoted by ``$+$'' and ``$\times$''. Let $\xi^+$ 
and $\xi^\times$ denote
independent random variables that characterize the statistics of the gravity
waves, their linear combinations will prove to be useful (ZS97)
\begin{eqnarray}
\nonumber
\xi^1 & = & (\xi^+ - i \xi^\times)/ \sqrt{2}, \\
\xi^2 & = & (\xi^+ + i \xi^\times)/ \sqrt{2},
\end{eqnarray}
which satisfy
\begin{eqnarray}
\nonumber
\langle \xi^{1*}({\bf k})\,\xi^1({\bf k}') \rangle &
    = & \langle \xi^{2*}({\bf k})\,\xi^2({\bf k}') \rangle
    = P^t(k)\,\delta({\bf k}-{\bf k}')/2, \\
\langle \xi^{1*}({\bf k})\,\xi^2({\bf k}') \rangle & = & 0,
\end{eqnarray}
where $P^t(k)$ is the primordial power spectrum for gravity waves.

For each Fourier mode ${\bf k}$, we choose $\hat{z} \parallel {\bf k}$; in this
frame, the temperature and polarization anisotropies
%[{\bf define the x-y!!!}],
due to the tensor perturbation are denoted as
$\Delta_T^t(\tau, {\bf k}, \hat{n})$ and $\Delta_P^t(\tau, {\bf k}, \hat{n}) \equiv
\Delta_Q^t + i \Delta_U^t$, where the $\Delta$ reminds us the contribution is from
one ${\bf k}$ mode, and the superscript ``t'' means ``tensor''. Note also, because
there is no azimuthal symmetry present now, the anisotropies depend on azimuthal
angle $\phi$ as well as $\mu \equiv \hat{k} \cdot \hat{n})$.
Their explicit form are \citep{pol85,k96}
\begin{eqnarray}
\nonumber
\Delta_T^t(\tau, {\bf k}, \hat{n}) & = & [(1-\mu^2)\,{\rm e}^{2i\phi}\,
  \xi^1({\bf k}) + (1-\mu^2)\,{\rm e}^{-2i\phi}\, \xi^2({\bf k})]\,
   \widetilde{\Delta}^t_T(\tau,k,\mu), \\
%(\Delta_Q^t + i \Delta_U^t)(\tau, {\bf k}, \hat{n}) & = &
\nonumber
\Delta_P^t(\tau, {\bf k}, \hat{n}) & = &
  [(1-\mu)^2\,{\rm e}^{2i\phi}\,\xi^1({\bf k}) +
  (1+\mu)^2\,{\rm e}^{-2i\phi}\,\xi^2({\bf k})]\,\widetilde{\Delta}^t_P(\tau,k,\mu), \\
\Delta_P^{t*}(\tau, {\bf k}, \hat{n}) & = &
  [(1+\mu)^2\,{\rm e}^{2i\phi}\,\xi^1({\bf k}) +
  (1-\mu)^2\,{\rm e}^{-2i\phi}\,\xi^2({\bf k})]\,\widetilde{\Delta}^t_P(\tau,k,\mu),
\end{eqnarray}
where the new variables $\widetilde{\Delta}^t_T(\tau,k,\mu)$ and
$\widetilde{\Delta}^t_P(\tau,k,\mu)$ are independent
of $\phi$ and satisfy the Boltzmann equation \citep{pol85}
\begin{eqnarray}
\nonumber
\dot{\widetilde{\Delta}^t_T} + ik\mu \widetilde{\Delta}^t_T & = &
  -\dot{h} -\dot{\kappa}\,(\widetilde{\Delta}^t_T -\Psi), \\
\nonumber
\dot{\widetilde{\Delta}^t_P} + ik\mu \widetilde{\Delta}^t_P & = &
   -\dot{\kappa}\,(\widetilde{\Delta}^t_P +\Psi), \\
\Psi & = & {1\over 10} \widetilde{\Delta}^t_{T0} + {1\over7}\widetilde{\Delta}^t_{T2}
  + {3\over70} \widetilde{\Delta}^t_{T4} - {3\over5} \widetilde{\Delta}^t_{P0}
  +{6\over7} \widetilde{\Delta}^t_{P2} -{3\over70} \widetilde{\Delta}^t_{P4}.
\end{eqnarray}

The solutions to these equations can be obtained by the line-of-sight
integration (ZS97):
\begin{eqnarray}
\nonumber
\Delta^t_T(\tau_0, {\bf k}, \hat{n}) & = &
  \left( (1-\mu^2)\,{\rm e}^{2i\phi}\,\xi^1({\bf k}) + (1-\mu^2)\,{\rm e}^{-2i\phi}
  \,\xi^2({\bf k}) \right)\,\int_0^{\tau_0}\,d\tau\,{\rm e}^{-ix\mu} S^t_T(\tau,k),\\
\nonumber
\Delta_P^t(\tau_0, {\bf k}, \hat{n}) & = &
  \left( (1-\mu)^2\,{\rm e}^{2i\phi}\,\xi^1({\bf k}) + (1+\mu)^2\,{\rm e}^{-2i\phi}
  \,\xi^2({\bf k}) \right) \,\int_0^{\tau_0}\,d\tau\,{\rm e}^{-ix\mu} S^t_P(\tau,k),\\
\Delta_P^{t*}(\tau_0, {\bf k}, \hat{n}) & = &
  \left( (1+\mu)^2\,{\rm e}^{2i\phi}\,\xi^1({\bf k}) + (1-\mu)^2\,{\rm e}^{-2i\phi}
  \,\xi^2({\bf k}) \right) \,\int_0^{\tau_0}\,d\tau\,{\rm e}^{-ix\mu} S^t_P(\tau,k),
\label{eq:tensorsoln}
\end{eqnarray}
where
\begin{eqnarray}
\nonumber
S^t_T(\tau,k) & = & -\dot{h}\,{\rm e}^{-\kappa}+g\,\Psi, \\
S^t_P(\tau,k) & = & -g\,\Psi.
\label{eq:tensorsource}
\end{eqnarray}

Next, to obtain spin--0 quantities we need to apply spin raising and lowering
operators to the integral solutions (ZS97):
\begin{eqnarray*}
\flat^2\, \Delta_P^t(\tau_0, {\bf k}, \hat{n}) & = &
  \xi^1({\bf k}) \int_0^{\tau_0}\,d\tau\,S^t_P(\tau,k) \left( -\partial_\mu +
  {2\over 1-\mu^2} \right)^2\,[(1-\mu^2)(1-\mu)^2\,{\rm e}^{2i\phi}{\rm e}^{-ix\mu}] \\
 & &  + \xi^2({\bf k}) \int_0^{\tau_0}\,d\tau\,S^t_P(\tau,k) \left( -\partial_\mu -
  {2\over 1-\mu^2} \right)^2\,[(1-\mu^2)(1+\mu)^2\,{\rm e}^{-2i\phi}{\rm e}^{-ix\mu}] \\
 & = & - \xi^1({\bf k})(1-\mu^2) \,{\rm e}^{2i\phi}\, \int_0^{\tau_0}\,d\tau\,S^t_P(\tau,k)
  \left( \hat{\mathcal{E}}(x) - i \hat{\mathcal{B}}(x) \right)\,{\rm e}^{-ix\mu} \\
 & &  - \xi^2({\bf k})(1-\mu^2) \,{\rm e}^{-2i\phi}\, \int_0^{\tau_0}\,d\tau\,S^t_P(\tau,k)
  \left( \hat{\mathcal{E}}(x) + i \hat{\mathcal{B}}(x) \right)\,{\rm e}^{-ix\mu}, \\
\sharp^2\, \Delta_P^{t*}(\tau_0, {\bf k}, \hat{n}) & = &
  \xi^1({\bf k}) \int_0^{\tau_0}\,d\tau\,S^t_P(\tau,k) \left( -\partial_\mu -
  {2\over 1-\mu^2} \right)^2\,[(1-\mu^2)(1+\mu)^2\,{\rm e}^{2i\phi}{\rm e}^{-ix\mu}] \\
 & &  + \xi^2({\bf k}) \int_0^{\tau_0}\,d\tau\,S^t_P(\tau,k) \left( -\partial_\mu +
  {2\over 1-\mu^2} \right)^2\,[(1-\mu^2)(1-\mu)^2\,{\rm e}^{-2i\phi}{\rm e}^{-ix\mu}] \\
 & = & - \xi^1({\bf k})(1-\mu^2) \,{\rm e}^{2i\phi}\, \int_0^{\tau_0}\,d\tau\,S^t_P(\tau,k)
  \left( \hat{\mathcal{E}}(x) + i \hat{\mathcal{B}}(x) \right)\,{\rm e}^{-ix\mu} \\
 & &  - \xi^2({\bf k})(1-\mu^2) \,{\rm e}^{-2i\phi}\, \int_0^{\tau_0}\,d\tau\,S^t_P(\tau,k)
  \left( \hat{\mathcal{E}}(x) - i \hat{\mathcal{B}}(x) \right)\,{\rm e}^{-ix\mu},
\end{eqnarray*}
where operators $\hat{\mathcal{E}}(x) \equiv -12 +x^2\,(1-\partial_x^2) -8x\,\partial_x$
and $\hat{\mathcal{B}}(x) \equiv 8x+2x^2\,\partial_x$ are introduced.
In obtaining these expressions, we have worked out the results of the
differentiations with respect to $\mu$, and used the trick
$f(b)\,{\rm exp}(-iab) = f(i \partial_a) \,{\rm exp}(-iab)$ again. For example,
the $\xi^1({\bf k})$ term of $\flat^2\, \Delta_P^t(\tau_0, {\bf k}, \hat{n})$:
\begin{eqnarray*}
\left( -\partial_\mu +{2\over 1-\mu^2} \right)^2\,
[(1-\mu^2)(1-\mu)^2\,{\rm e}^{-ix\mu}]\! & = &\!
  -(1-\mu^2) \left( -12-8i(1-\mu)x+(1-\mu)^2 x^2 \right) {\rm e}^{-ix\mu} \\
\!& = & \!-(1-\mu^2) \left(-12-8ix-8x\partial_x+x^2(1-i\partial_x)^2\right) {\rm e}^{-ix\mu} \\
\!& = & \!-\left(\hat{\mathcal{E}}(x) - i \hat{\mathcal{B}}(x) \right)\,(1-\mu^2){\rm e}^{-ix\mu}.
\end{eqnarray*}
Notice how these operators simplify the $\mu$ dependence of the
expressions. From these we construct
$\Delta_{\widetilde{E}}^t(\tau_0,{\bf k},\hat{n})$ and
$\Delta_{\widetilde{B}}^t(\tau_0,{\bf k},\hat{n})$:
\begin{eqnarray}
\nonumber
\Delta^t_{\widetilde{E}}(\tau_0,{\bf k}, \hat{n}) & = & {-1\over2}
        \left( \flat^2 \Delta_P^t(\tau_0, {\bf k}, \hat{n})
        + \sharp^2 \Delta_P^{t*}(\tau_0, {\bf k}, \hat{n}) \right) \\
\label{eq:deltaet}
 & = &
  (1-\mu^2) \left({\rm e}^{2i\phi}\,\xi^1({\bf k}) + {\rm e}^{-2i\phi}
  \,\xi^2({\bf k}) \right) \hat{\mathcal{E}}(x)
  \,\int_0^{\tau_0}\,d\tau\,{\rm e}^{-ix\mu} S^t_P(\tau,k), \\
\nonumber
\Delta^t_{\widetilde{B}}(\tau_0,{\bf k}, \hat{n}) & = & {-1\over2i}
        \left( \flat^2 \Delta_P^t(\tau_0, {\bf k}, \hat{n})
        - \sharp^2 \Delta_P^{t*}(\tau_0, {\bf k}, \hat{n}) \right) \\
\label{eq:deltabt}
 & = &
  -(1-\mu^2) \left({\rm e}^{2i\phi}\,\xi^1({\bf k}) - {\rm e}^{-2i\phi}
  \,\xi^2({\bf k}) \right) \hat{\mathcal{B}}(x)
  \,\int_0^{\tau_0}\,d\tau\,{\rm e}^{-ix\mu} S^t_P(\tau,k).
\end{eqnarray}
The total contributions from all ${\bf k}$ modes are then
\begin{equation}
X^t(\hat{n}) = \int d^3{\bf k}\,\Delta_X^t(\tau_0, {\bf k}, \hat{n}),
\end{equation}
where $X = T,\,\widetilde{E},\,\widetilde{B}$. The solution
$\Delta^t_T(\tau_0, {\bf k}, \hat{n})$ is already given in
Eqn~(\ref{eq:tensorsoln}). These are our starting points in calculating the
power spectra.

First let us calculate the temperature spectrum:
\begin{eqnarray*}
C_{Tl} & = & {1\over 2l+1} \sum_m \langle a_{T,lm}^*\,a_{T,lm} \rangle \\
 & = & {1\over 2l+1} \sum_m \langle
  \left( \int\,d\Omega'\,Y_{lm}^*(\hat{n}')\,T^t(\hat{n}') \right)^*
  \left( \int\,d\Omega\,Y_{lm}^*(\hat{n})\,T^t(\hat{n}) \right) \rangle \\
 & = & {1\over 2l+1} \sum_m \langle \int\,d\Omega'\,d\Omega\,d^3{\bf k}'\,d^3{\bf k}
  \,Y_{lm}(\hat{n}')\,Y_{lm}^*(\hat{n})\,
  \Delta^{t*}_T(\tau_0,k',\mu')\,\Delta^t_T(\tau_0,k,\mu) \rangle \\
 & = & {1\over 2l+1} \sum_m \langle \int\,d\Omega'\,d\Omega\,d^3{\bf k}'\,d^3{\bf k}
  \,Y_{lm}(\hat{n}')\,Y_{lm}^*(\hat{n}) \\
 & & \mbox{\ \ \ } \times (1-\mu'^2)
  ({\rm e}^{-2i\phi'}\,\xi^{1*}({\bf k}')+{\rm e}^{2i\phi'}\,\xi^{2*}({\bf k}'))
  \left( \int_0^{\tau_0} d\tau\,{\rm e}^{-ix\mu'}\,S_T^t(\tau,k') \right)^* \\
 & & \mbox{\ \ \ } \times \,(1-\mu^2)
  ({\rm e}^{2i\phi}\,\xi^{1}({\bf k})+{\rm e}^{-2i\phi}\,\xi^{2}({\bf k}))
  \int_0^{\tau_0} d\tau\,{\rm e}^{-ix\mu}\,S_T^t(\tau,k) \rangle \\
 & = & {1\over 2l+1} \sum_m \int\,d\Omega'\,d\Omega\,d^3{\bf k}'\,d^3{\bf k}
  \,Y_{lm}(\hat{n}')\,Y_{lm}^*(\hat{n}) \\
 & & \mbox{\ \ \ } \times (1-\mu'^2)(1-\mu^2)\,
  \left( \int_0^{\tau_0} d\tau\,{\rm e}^{-ix\mu'}\,S_T^t(\tau,k') \right)^*\,
  \int_0^{\tau_0} d\tau\,{\rm e}^{-ix\mu}\,S_T^t(\tau,k) \\
 & &  \mbox{\ \ \ } \times
  ({\rm e}^{2i\phi} {\rm e}^{-2i\phi'}+{\rm e}^{-2i\phi} {\rm e}^{2i\phi'})
  \,{1\over2}\,P^t(k)\,\delta({\bf k}-{\bf k}') \\
 & = & {1\over 2l+1} \sum_m \int\,d^3{\bf k}\,d\Omega'\,d\Omega
  \,Y_{lm}(\hat{n}')\,Y_{lm}^*(\hat{n})
  (1-(\hat{k} \cdot \hat{n})^2)(1-(\hat{k} \cdot \hat{n}')^2) \\
 & & \mbox{\ \ \ } \times
  \left( \int_0^{\tau_0} d\tau\,{\rm e}^{-ix \hat{k} \cdot \hat{n}'}\,S_T^t(\tau,k) \right)^* \,
  \left( \int_0^{\tau_0} d\tau\,{\rm e}^{-ix \hat{k} \cdot \hat{n}}\,S_T^t(\tau,k) \right)\,
  {\rm e}^{2i\phi} {\rm e}^{-2i\phi'} \,{1\over2}\,P^t(k) \\
 & &  + {1\over 2l+1} \sum_m \int\,d^3{\bf k}\,d\Omega'\,d\Omega
  \,Y_{lm}(\hat{n}')\,Y_{lm}^*(\hat{n})
  (1-(\hat{k} \cdot \hat{n})^2)(1-(\hat{k} \cdot \hat{n}')^2) \\
 & & \mbox{\ \ \ } \times
  \left( \int_0^{\tau_0} d\tau\,{\rm e}^{-ix \hat{k} \cdot \hat{n}'}\,S_T^t(\tau,k) \right)^*\,
  \left( \int_0^{\tau_0} d\tau\,{\rm e}^{-ix \hat{k} \cdot \hat{n}}\,S_T^t(\tau,k) \right)\,
  {\rm e}^{-2i\phi} {\rm e}^{2i\phi'} \,{1\over2}\,P^t(k);
\end{eqnarray*}
we will prove the two terms on the right hand side are equal. 
Let us denote the first term as $t_I$, the second term $t_{II}$.
The following equations and expressions will be used:
\[
\int_0^{2\pi}\,d\phi\, {\rm e}^{\pm 2i\phi}\,{\rm e}^{-im\phi} =
2\pi \delta_{m\pm 2},
\]
the relation between $Y_{lm}$ and associated Legendre polynomials 
\citep[e.g.][]{jackson}
\[
Y_{lm}(\mu,\phi) = \left[{2l+1\over 4\pi} {(l-m)! \over (l+m)!} \right]^{1/2}
\,P_{lm}(\mu)\,{\rm e}^{im\phi},
\]
the explicit form of associated Legendre polynomials
\[
P_{lm}(\mu) = (-1)^m (1-\mu^2)^{m/2} {d^m \over d\mu^m} P_l(\mu),
\]
and the relation between $P_{lm}(\mu)$ and $P_{l-m}(\mu)$
\[
P_{l-m}(\mu) = (-1)^m {(l-m)! \over (l+m)!} P_{lm}(\mu).
\]

We begin with the term $t_I$:
\begin{eqnarray*}
t_I & = & {2\pi\over 2l+1} \sum_m \int\,k^2\,dk\,P^t(k)\,\left| \int d\Omega
  \,Y_{lm}^*(\hat{n})\,(1-\mu^2)\,{\rm e}^{2i\phi}\,
  \int_0^{\tau_0} d\tau\,{\rm e}^{-ix\mu}\,S_T^t(\tau,k) \right|^2 \\
 & = & {2\pi\over 2l+1} \int\,k^2\,dk\,P^t(k) \\
 & & \mbox{\ \ \ } \times \sum_m \,\left| \int\,d\phi\,
  {\rm e}^{i(2-m)\phi}\,\int d\mu
  \left[{2l+1\over 4\pi} {(l-m)! \over (l+m)!} \right]^{1/2}\,P_{lm}(\mu)
  \,(1-\mu^2)\,\int_0^{\tau_0} d\tau\,{\rm e}^{-ix\mu}\,S_T^t(\tau,k) \right|^2 \\
 & = & 2\pi^2 \,{(l-2)! \over (l+2)!} \int\,k^2\,dk\,P^t(k)\,
  \left| \int d\mu (1-\mu^2)\,P_{l2}(\mu)\,\int_0^{\tau_0} d\tau\,
  {\rm e}^{-ix\mu}\,S_T^t(\tau,k) \right|^2,
\end{eqnarray*}
the second term $t_{II}$ is:
\begin{eqnarray*}
t_{II} & = & {2\pi\over 2l+1} \sum_m \int\,k^2\,dk\,P^t(k)\,\left| \int d\Omega
  \,Y_{lm}^*(\hat{n})\,(1-\mu^2)\,{\rm e}^{-2i\phi}\,
  \int_0^{\tau_0} d\tau\,{\rm e}^{-ix\mu}\,S_T^t(\tau,k) \right|^2 \\
 & = & {2\pi\over 2l+1} \int\,k^2\,dk\,P^t(k) \\
 & & \mbox{\ \ } \times \sum_m \,\left| \int\,d\phi\,
  {\rm e}^{i(-2-m)\phi} \int d\mu
  \left[{2l+1\over 4\pi} {(l-m)! \over (l+m)!} \right]^{1/2} P_{lm}(\mu)
  \,(1-\mu^2)\!\int_0^{\tau_0} d\tau\,{\rm e}^{-ix\mu}\,S_T^t(\tau,k) \right|^2 \\
 & = & 2\pi^2 \,{(l+2)! \over (l-2)!} \int\,k^2\,dk\,P^t(k)\,
  \left| \int d\mu (1-\mu^2)\,\underbrace{P_{l-2}(\mu)}_{P_{l2} (l-2)!/(l+2)!}
  \,\int_0^{\tau_0} d\tau\, {\rm e}^{-ix\mu}\,S_T^t(\tau,k) \right|^2 \\
 & = & t_I 
  =  2\pi^2 \,{(l-2)! \over (l+2)!} \int\,k^2\,dk\,P^t(k)\,
  \left| \int_0^{\tau_0} d\tau\, S_T^t(\tau,k)\,\int_{-1}^1 d\mu\,
  (1-\mu^2)\,P_{l2}(\mu)\,{\rm e}^{-ix\mu} \right|^2.
\end{eqnarray*}
Using the definition of associated Legendre polynomials,
the angular integration can be written as
\begin{eqnarray*}
\int_{-1}^1\,d\mu\,(1-\mu^2)\,P_{l2}(\mu)\,{\rm e}^{-ix\mu} & = &
  \int_{-1}^1\,d\mu\,\left( {d^2 \over d\mu^2} P_l(\mu)\right)
  (1-\mu^2)^2\,{\rm e}^{-ix\mu} \\
 & = & \int_{-1}^1\,d\mu\,\left( {d^2 \over d\mu^2} P_l(\mu)\right)
  (1+\partial_x^2)^2 {\rm e}^{-ix\mu} \\
% & = & \int_{-1}^1\,d \left( {dP_l \over d\mu} \right)
%  (1+\partial_x^2)^2 {\rm e}^{-ix\mu} \\
 & = & \left. {dP_l \over d\mu} (1+\partial_x^2)^2
  {\rm e}^{-ix\mu} \right|^1_{-1} - \int_{-1}^1\,d\mu\,{dP_l \over d\mu}
  \,(1+\partial_x^2)^2 (-ix{\rm e}^{-ix\mu}) \\
 & = & \left. {dP_l \over d\mu} {\rm e}^{-ix\mu}(1-\mu^2)^2\right|^1_{-1}
  + \left. P_l(\mu)\,(1+\partial_x^2)^2 (ix{\rm e}^{-ix\mu}) \right|^1_{-1} \\
 & & \mbox{\ \ \ }
  -\int_{-1}^1\,d\mu\,P_l(\mu)(1+\partial_x^2)^2(x^2{\rm e}^{-ix\mu}) \\
 & = & 0 + \left. P_l(\mu)\,i (4i\mu-x+\mu^2 x) \,{\rm e}^{-ix\mu}(1-\mu^2)^2
  \right|^1_{-1} \\
 & & \mbox{\ \ \ } - (1+\partial_x^2)^2 \left( x^2 \int_{-1}^1 d\mu\,
  P_l(\mu)\,{\rm e}^{-ix\mu} \right) \\
 & = & -2 (1+\partial_x^2)^2 (x^2(-i)^l\,j_l(x)),
\end{eqnarray*}
where we have used
\[
\int_{-1}^1 d\mu\,{\rm e}^{-ix\mu} P_l(\mu) = 2\,(-i)^l\,j_l(x).
\]

Therefore the temperature power spectrum is (ZS97):
\begin{eqnarray}
\nonumber
C_{Tl} & = & 2\,t_I \\
\nonumber
 & = & 4\pi^2 \,{(l-2)! \over (l+2)!} \int\,k^2\,dk\,P^t(k)\,
  \left| \int_0^{\tau_0} d\tau\, S_T^t(\tau,k)\,
  2 (1+\partial_x^2)^2 (x^2 (-i)^l\,j_l(x)) \right|^2 \\
\nonumber
 & = & (4\pi)^2\,{(l-2)! \over (l+2)!} \int\,k^2\,dk\,P^t(k)\,
  \left| \int_0^{\tau_0} d\tau\, S_T^t(\tau,k)\,
  {j_l(x) \over x^2} \right|^2 \\
 & = & (4\pi)^2 \int k^2\,dk\,P^t(k) [\Delta^t_{Tl}(k)]^2,
\end{eqnarray}
where we have used the expression Eqn~(\ref{eq:j}), the same trick
used in calculation of scalar polarization power spectrum, and defined
\begin{equation}
\Delta^t_{Tl}(k) \equiv \left[{(l-2)! \over (l+2)!} \right]^{1/2}
  \int_0^{\tau_0}d\tau \,S_T^t(\tau,k)\,{j_l(x) \over x^2},
\label{eq:deltattensor}
\end{equation}
where $S_T^t(\tau,k)$ is given in Eqn~(\ref{eq:tensorsource}).
Notice the similarities between these expressions and
Eqns~(\ref{eq:ctlscalar}) \& (\ref{eq:ctlscalarsource}).

This formulation has a big payoff, because the calculations of
remaining spectra are very similar and straightforward. The reason is that 
the angular dependence of
$\Delta^t_{\widetilde{E}}(\tau_0, {\bf k}, \hat{n})$ and
$\Delta^t_{\widetilde{B}}(\tau_0, {\bf k}, \hat{n})$ are the same as
that of $\Delta^t_{E}(\tau_0, {\bf k}, \hat{n})$. 
This is clearly shown in Eqns~(\ref{eq:tensorsoln}), (\ref{eq:deltaet}) \& (\ref{eq:deltabt}). The difference,
namely the operators $\hat{\mathcal{E}}$ and $\hat{\mathcal{B}}$,
can be applied after the angular integrations are performed (ZS97).
This leads to
\begin{eqnarray}
\nonumber
C_{El} & = & (4\pi)^2\,\int\,k^2\,dk\,P^t(k)\,
  \left| \int_0^{\tau_0} d\tau\, S_P^t(\tau,k)\,\hat{\mathcal{E}}\,
  {j_l(x) \over x^2} \right|^2 \\
\label{eq:cel}
 & = & (4\pi)^2\,\int\,k^2\,dk\,P^t(k)\,\left| \int_0^{\tau_0}
  d\tau\, S_P^t(\tau,k)\, \left[ -j_l(x)+j_l''(x) +{2j_l(x) \over x^2}
  + {4j_l'(x) \over x} \right] \right|^2, \\
\nonumber
C_{Bl} & = & (4\pi)^2\,\int\,k^2\,dk\,P^t(k)\,
  \left| \int_0^{\tau_0} d\tau\, S_P^t(\tau,k)\,\hat{\mathcal{B}}\,
  {j_l(x) \over x^2} \right|^2 \\
\label{eq:cbl}
 & = & (4\pi)^2\,\int\,k^2\,dk\,P^t(k)\,\left| \int_0^{\tau_0}
  d\tau\, S_P^t(\tau,k)\, \left[ 2j_l'(x) +{4j_l(x) \over x}
  \right] \right|^2.
\end{eqnarray}

These equations can be further simplified by integrating by parts the $j_l'$ and
$j_l''$ terms to increase computational efficiency (ZS97). The source term 
$S_P^t(\tau,k) = -g \Psi$ (Eqn \ref{eq:tensorsource}) makes sure the boundary 
terms at $\tau = 0$ vanish; the boundary terms at $\tau=\tau_0$ can be
ignored for $l>0$ modes (i.e. excluding the monopole; see the
Appendix B). The $j_l''$ term in the integrand of $C_{El}$
gives $-j_l(x) (\ddot{g} \Psi + 2\dot{g} \dot{\Psi} + 
g \ddot{\Psi})/k^2$, the $4j_l'/x$ term gives $-4j_l(x) (
%\dot{g} \Psi/kx + g \dot{\Psi}/kx - g\Psi/x^2)$; the $2j_l'$
\dot{g} \Psi/kx + g \dot{\Psi}/kx + g\Psi/x^2)$; the $2j_l'$
term in the integrand of $C_{Bl}$ gives $-2j_l(x)(\dot{g}\Psi
+g\dot{\Psi})/k$.
The final expressions are (ZS97)
\begin{eqnarray}
\nonumber
C_{Xl} & = & (4\pi)^2 \int k^2\,dk\,P^t(k)\, [\Delta^t_{Xl}(k)]^2, \\
C_{Cl} & = & (4\pi)^2 \int k^2\,dk\,P^t(k)\, \Delta^t_{Tl}(k)\,\Delta^t_{El}(k),
\end{eqnarray}
where $X = T,\,E,\,B$; $\Delta_{Tl}^t(k)$ is given in
Eqn~(\ref{eq:deltattensor}), and
\begin{eqnarray}
\nonumber
\Delta^t_{(E,B)l}(k) & = & \int_0^{\tau_0}\,d\tau\,S^t_{(E,B)}(\tau,k)\,j_l(x),\\
\nonumber
S^t_{E}(\tau,k) & = & g(\tau) \left( -\Psi + { \ddot{\Psi} \over k^2}
%  +{2\Psi \over x^2} -{4\dot{\Psi} \over kx} \right) -\dot{g}(\tau) \left(
  +{6\Psi \over x^2} +{4\dot{\Psi} \over kx} \right) +\dot{g}(\tau) \left(
  {2\dot{\Psi} \over k^2} + {4\Psi \over kx} \right) +\ddot{g}(\tau){ \Psi \over k^2}, \\
S^t_{B}(\tau,k) & = & g(\tau)\left( {4\Psi \over x} + {2\dot{\Psi} \over k} \right)
  +2\dot{g}(\tau) {\Psi \over k}.
\end{eqnarray}

\section{Summary}

We have given a detailed discussion of the physics of polarization in the
CMB. Like the temperature anisotropy, polarization is anisotropic over the
sky. After a review of the mathematical and physical background (\S~2), 
we have re-derived  the statistical properties of the temperature and polarization
anisotropy in terms of the power spectra in the line-of-sight formalism (\S~3,
see also Appendices A and B), which are implemented in the code CMBFAST (ZS97). 
We hope that this uniform treatment will be of value for the beginning 
theorist learning about polarization and the solution of the linearized
cosmological Boltzmann equations. We provide some suggested reading material in
the next section.

\section{Further Reading}

In this review we follow the mathematical formalism developed in \citet[][ZS97]{zs97} to
construct polarization power spectra. Our treatment can be complemented by the
article by \citet{hu97a}, who provided vivid visual aids to understanding the physics of
polarization\footnote{see http://background.uchicago.edu/$\sim$whu/polar/webversion/polar.html}.  A recent review \citep{cabella04} treated the same
subject under the tensor spherical harmonics formalism \citep{kks97}.

The technique presented in ZS97 was for the full-sky in a spatially flat 
Universe (which is strongly favored by many observations, especially those
from the {\it WMAP} satellite). 
\citet{zsb98} extended the integral solutions and the calculation of
polarization field to non--flat Universe models. \citet{hu97b,hu98} unified the
treatments for anisotropies produced by scalar,
vector and tensor perturbations. If this guide is to be for beginners, we 
refer our readers to these treatments for the intermediate level.
A quantum mechanically oriented discussion can be found in \citet{k96}.

For more details on simulating the CMB polarization maps and their
statistical properties, we refer to the discussions in ZS97 (\S~V) and
\citet{kks97}.  Detailed and general discussions of polarization
experiments can be found in, e.g., \citet{sel97,hu97b,z98b,teg01,dasi,wmappol,
bond03,bunn03} and references therein.
For more specific discussions on individual polarization experiments (completed 
or on-going), see e.g.
\citet{keating01,hedman01,dasi,johnson03,montroy03,keating03,deoliverira03,barkats04,farese04,bowden04,cortiglioni04,leitch04,readhead04}.

General discussions of the importance of polarization in constraining or 
breaking degeneracies in 
cosmological parameters, or helping to discover new physics can be found in,
e.g. \citet{zss97,k99,eisenstein99,teg04} and the references therein. For more
specific topics, such as the reionization and polarization, see e.g.
\citet{hu00d,liu01,haiman03}; for weak gravitational lensing and polarization, see e.g. 
\citet{zaldarriaga98d,stompor99,kesden02,seljak04}.

For recent general review on CMB and its polarization, see, e.g. \citet{k01,
hu02,z03,challinor04}. We also recommend the treatments in textbooks by 
\citet{peacock,ll,dodelson} on this topic.

Among the many useful websites that provide up-to-date information or detailed
exposures on CMB-related topics, we refer the reader to that of Wayne 
Hu\footnote{http://background.uchicago.edu/$\sim$whu/index.html} for CMB tutorials,
that of Anthony Banday\footnote{http://www.mpa-garching.mpg.de/$\sim$banday/CMB.html} for a detailed listing of CMB resources, that of Max 
Tegmark\footnote{http://www.hep.upenn.edu/$\sim$max/} for CMB data analysis, those of Martin White\footnote{http://astron.berkeley.edu/$\sim$mwhite/htmlpapers.html} and
Ned Wright\footnote{http://www.astro.ucla.edu/$\sim$wright/intro.html} for
cosmology tutorials, and Edmund Bertschinger's pages\footnote{http://arcturus.mit.edu/$\sim$edbert/} for excellent lecture notes on cosmology and structure formation. An online reading list that also provides links to many additional useful
websites can be found on Martin White's page\footnote{http://astron.berkeley.edu/$\sim$mwhite/readinglist.html}.

Finally, for numerical work we recommend
HEALPix\footnote{http://www.eso.org/science/healpix/}, publically available
mathematical software for the simulation and analysis of temperature and 
polarization maps on the sphere \citep{ghw,gbhw}.

\acknowledgements
We thank an anonymous referee for suggestions and corrections that improve the 
presentation of 
the manuscript, and David Larson for helpful comments. YTL thanks
Jimmy Snyder for comments on an earlier version of the manuscript.
This work has been partially supported by the University of Illinois at Urbana-Champaign.

%%%%%%%%%%%%%%%%%%%%%%%%%%%%%%%%%%%%%%%%%%%%%%%%%%
%%%%%%%%%%%%%%%%%%%%%%%%%%%%%%%%%%%%%%%%%%%%%%%%%%
%%%%%%%%%%%%%%%%%%%%%%%%%%%%%%%%%%%%%%%%%%%%%%%%%%

\appendix

\section{Parity of E and B Modes
\label{sec:parity}}

Here we discuss the properties of Stokes $Q$ and $U$ and
$\widetilde{E}$ and $\widetilde{B}$ (namely $E$ and $B$) under parity transformations.
The discussion is similar to that of \citet{z98}.
Let us consider the space inversion $r \rightarrow r$, $\theta \rightarrow
\pi-\theta$, $\phi \rightarrow \pi + \phi$. 
In the language of basis kets used in \S~\ref{sec:qm}, a polarization state is
characterized by the expectation value of the Stokes operators. If the 
the basis kets are chosen as $|\epsilon_1\rangle = \hat{e}_\theta$, 
$|\epsilon_2\rangle = \hat{e}_\phi$, then they are related to their counterparts
in the space--inversed frame by $|\epsilon_1'\rangle = \hat{e}_\theta' 
= -|\epsilon_1\rangle $ and $|\epsilon_2'\rangle = \hat{e}_\phi' = 
|\epsilon_2\rangle $. We therefore see that the expectation values are $Q' = Q$
and $U' = -U$: $Q$ has even parity, while $U$ has odd parity.

As for the parities of $E$ and $B$, it is useful to notice that $\partial_{\phi'}
 = \partial_\phi$, $\partial_{\theta'} = -\partial_\theta$.
Let $\hat{n}$ and $\hat{n}'$ refer to the same physical direction in
the original and space--inversed frames, respectively.
From $(Q+iU)'(\hat{n}') = (Q-iU)(\hat{n})$ we have, using
Eqn~(\ref{eq:sandf}),
\begin{eqnarray*}
\flat' (Q+iU)'(\hat{n}') & = & -{1\over \sin^2\theta'} \left(
  \partial_{\theta'} -{i\over \sin\theta'} \partial_{\phi'} \right)
  \sin^2\theta'\,(Q+iU)'(\hat{n}') \\
 & = & {1\over \sin^2\theta} \left(\partial_{\theta}
  +{i\over \sin\theta}\partial_{\phi} \right)\sin^2\theta\,(Q-iU)(\hat{n}) \\
 & = & -\sharp (Q-iU)(\hat{n}),  \\
\flat'^2 (Q+iU)'(\hat{n}') & = & -{1\over \sin\theta'} \left(
  \partial_{\theta'} -{i\over \sin\theta'} \partial_{\phi'} \right)
  \sin\theta'\,\flat' (Q+iU)'(\hat{n}') \\
 & = & {1\over \sin\theta} \left(\partial_{\theta}
  +{i\over \sin\theta}\partial_{\phi} \right)\sin\theta\,\left(-\sharp\,(Q-iU)(\hat{n}) \right) \\
 & = & \sharp^2 (Q-iU)(\hat{n}).
\end{eqnarray*}
Similarly, we have $\sharp'^2 (Q+iU)'(\hat{n}') = \flat^2 (Q-iU)(\hat{n})$.

Using these results and the definition of the $\widetilde{E}$ and $\widetilde{B}$ modes,
\begin{eqnarray*}
-2 \widetilde{E'}(\hat{n}') & = & \flat^2 (Q+iU)(\hat{n}') + \sharp^2 (Q-iU)(\hat{n}'), \\
-2i \widetilde{B}(\hat{n}') & = & \flat^2 (Q+iU)(\hat{n}') - \sharp^2 (Q-iU)(\hat{n}'),
\end{eqnarray*}
we see that
\begin{eqnarray*}
-2 \widetilde{E'}(\hat{n}') & = & \sharp^2 (Q-iU)(\hat{n}) + \flat^2 (Q+iU)(\hat{n}) = -2 \widetilde{E}(\hat{n}), \\
-2i \widetilde{B'}(\hat{n}') & = & \sharp^2 (Q-iU)(\hat{n}) - \flat^2 (Q+iU)(\hat{n}) = 2i\widetilde{B}(\hat{n}).
\end{eqnarray*}
Therefore $E$ and $\widetilde{E}$ have even parities, while $B$ and $\widetilde{B}$ have 
odd parities.

%%%%%%%%%%%%%%%%%%%%%%%%%%%%%%%%%%%%%%%%%%%%%%%%%%
\section{Line-of-Sight Integral Solution to Boltzmann Equation
\label{sec:soln}}

In this section we discuss the integral solution to the Boltzmann equation
\citep{sz96}. We first consider the polarization due to scalar perturbations.
The Boltzmann equation is
\begin{eqnarray*}
%\dot{\Delta}_T^s + ik\mu \Delta_T^s & = & -{1\over6}\dot{h}
%  -{1\over6}(\dot{h}+6\dot{\eta}) P_2(\mu) \\
% & & +\dot{\kappa} \left[ -\Delta_T^s + \Delta_{T0}^s + i\mu v_b
%     +{1\over2} P_2(\mu) \Pi \right], \\
\dot{\Delta}_P + ik\mu \Delta_P & = & \dot{\kappa} \left[ -\Delta_P
        + {3\over4}(1-\mu^2) \Pi \right], \\
\Pi(\tau,k) & = & \Delta_{T2} +\Delta_{P0} +\Delta_{P2},
\end{eqnarray*}
where, for brevity, the superscript $s$ is omitted.
If we move the term $-\dot{\kappa}\Delta_P$ on the right--hand side
to the other side of equality, and multiply both sides with
${\rm e}^{ik\mu\tau-\kappa}$, we find the whole expression becomes
${d\over d\tau}(\Delta_P\,{\rm e}^{ik\mu\tau-\kappa})$. (Because
of the way it is defined, 
$d\,{\rm e}^{-\kappa}/d\tau = \dot{\kappa}\,
{\rm e}^{-\kappa}$; c.f. Eqn~\ref{eq:kappa}.)  Integrating over the conformal time then gives
\begin{eqnarray*}
\int_0^{\tau_0}\,d\tau\,{d\over d\tau}\,
  \left(\Delta_P\,{\rm e}^{ik\mu\tau-\kappa} \right) & = &
  \Delta_P\,{\rm e}^{ik\mu\tau_0} - \Delta_P\,{\rm e}^{-\kappa(\tau=0)} \\
 & = & {3\over4}\,\int_0^{\tau_0}\,d\tau\,{\rm e}^{ik\mu\tau-\kappa}\,
  \dot{\kappa}(1-\mu^2)\Pi,
\end{eqnarray*}
where, by definition, $\kappa(\tau_0) = 0$, ${\rm e}^{-\kappa(\tau_0)} = 
1$, and we have used the fact that $\kappa(\tau=0) \rightarrow \infty$
(i.e. the optical depth to the ``big bang'' is infinite), the
$\Delta_P\,{\rm e}^{-\kappa(\tau=0)}$ term vanishes; after dividing through
the factor ${\rm e}^{ik\mu\tau_0}$, we have
\begin{eqnarray*}
\Delta_P(\tau_0,k,\mu) & = & {3\over4}(1-\mu^2)\, \int_0^{\tau_0}\,d\tau\,
  {\rm e}^{-ix\mu}\,\dot{\kappa}{\rm e}^{-\kappa}\,\Pi,
\end{eqnarray*}
the solution (\ref{eq:bsoln}), where $x = k(\tau_0-\tau)$. 

The solution to the temperature anisotropies is obtained analogously. 
%See \citet{sz96}
%for more details (note they worked in the longitudinal gauge in that paper).
We have
\begin{eqnarray*}
\Delta_T(\tau_0,k,\mu) & = & \int_0^{\tau_0}\,d\tau\,
  {\rm e}^{-ix\mu}\,{\rm e}^{-\kappa}\,\left[ -{1\over6}\dot{h}
  -{1\over3}(\dot{h}+6\dot{\eta}) P_2(\mu) 
 +\dot{\kappa} \left( \Delta_{T0}^s - i\mu v_b
     -{1\over2} P_2(\mu) \Pi \right) \right].
\end{eqnarray*}
We can further simplify the $\mu$ dependences in the integral by
taking the advantage of the ${\rm e}^{-ix\mu}$ term: for an arbitrary
function $Y$ of $\tau$ and $\mu$, we have
\begin{eqnarray*}
\int_0^{\tau_0}\!\! d\tau {\rm e}^{-ix\mu} {\rm e}^{-\kappa} Y(\tau,\mu)\! & = &
  \!\!\int_0^{\tau_0} d\tau\,{1 \over i k \mu}\,{d {\rm e}^{-ix\mu} \over d \tau} 
  {\rm e}^{-\kappa}\,Y(\tau,\mu) \\
%  & = & {1 \over i k \mu} \left\{ {\rm e}^{ix\mu} \left. 
%  {\rm e}^{-\kappa}\,Y(\tau) \right|_{0}^{\tau_0} -\int d\tau {\rm e}^{ix\mu}
%  {d \over d\tau} \left( {\rm e}^{-\kappa}\,Y \right) \right\} \\
  \!& = & \!\!{1 \over i k \mu} \left\{ {\rm e}^{-\kappa(\tau_0)} Y(\tau_0,\mu)
  - {\rm e}^{-i\mu k \tau_0- \kappa(0)} Y(0,\mu) 
  - \!\int_0^{\tau_0}\! d\tau {\rm e}^{-ix\mu} {d \over d\tau} \left( {\rm e}^{-\kappa}\,Y \right) \right\},
\end{eqnarray*}
the first term on the right-hand side is simply $Y(\tau_0)$, which can be
regarded as a constant contribution to the photon temperature (the 
``monopole''), and can be ignored since we are only interested in the 
temperature fluctuations; the second term vanishes because of the damping
before the last scattering. Therefore, if $Y(\tau,\mu)$ contains integral powers
of $\mu$, we can eliminate its $\mu$ dependences by successive 
applications of the above identity. To proceed, let us group the terms
on the right-hand side of the integral solution by powers of $\mu$.
Recall that $\alpha = (\dot{h}+6\dot{\eta})/2k^2$, the first
two terms in the square brackets reduce to 
$\dot{\eta}-k^2\mu^2\alpha$. We therefore have
\begin{eqnarray*}
\Delta_T(\tau_0,k,\mu) & = & \int_0^{\tau_0}\,d\tau\,
  {\rm e}^{-ix\mu}\,\left[ \left( {\rm e}^{-\kappa}\dot{\eta} + g \Delta_{T0}
  +  {1 \over 4}g \Pi \right) -i v_b g \,\mu
  - \left( k^2 \alpha\, {\rm e}^{-\kappa} + {3 \over 4}g \Pi  \right) \mu^2 \right],
\end{eqnarray*}
where we have used $g = \dot{\kappa} {\rm e}^{-\kappa}$. Integrating by
parts, the term proportional to $\mu$ is replaced by $(\dot{v}_b g + v_b \dot{g})/k$,
while the terms proportional to $\mu^2$ give
\[
\left( \ddot{\alpha} {\rm e}^{-\kappa} + 2 \dot{\alpha} g + \alpha \dot{g} \right)
  + {3\over 4 k^2} \left( g\ddot{\Pi}  + 2 \dot{g} \dot{\Pi} + \ddot{g} \Pi \right).
\]
Grouping these in derivatives with respect to $g$, we recover the source
term $S_T^s(\tau,k)$ in Eqn~(\ref{eq:bsoln}).

%%%%%%%%%%%%%%%%%%%%%%%%%%%%%%%
%\newpage

\end{document}